\documentclass[a4paper, notoc]{JHEP3}
\usepackage[dvips]{graphicx}
\usepackage{amsfonts}
\usepackage{amssymb}
\usepackage{epsfig}
\usepackage{cite}
\usepackage{multicol}
\usepackage{axodraw}
\usepackage{subfigure}
\usepackage{colordvi}
\DeclareGraphicsExtensions{.eps}
\usepackage{amsmath}
\usepackage{graphics,subfigure}
\usepackage{float}
\usepackage{booktabs}
\usepackage{rotating}

\def\eq{\begin{equation}}
\def\qe{\end{equation}}
\def\eqa{\begin{eqnarray}}
\def\qea{\end{eqnarray}}

\newcommand{\newc}{\newcommand}

\newc{\softsusy}{\texttt{SOFTSUSY}}
\newc{\isajet}{\texttt{ISAJET 7.64}}
\newc{\wigisasusy}{\texttt{WIGISASUSY}}

\newc{\ifb}{\textrm{fb}^{-1}}

\newc{\MO}{M_0}
\newc{\Mhalf}{M_{1/2}}
\newc{\AO}{A_0}
\newc{\tanb}{\textrm{tan}\beta}
\newc{\sgnmu}{\textrm{sgn}{\mu}}

\newc{\bbdecay}{0\nu\beta\beta}
\newc{\betadecay}{0\nu\beta\beta}
\newc{\bhalflife}{T^{0\nu\beta\beta}_{1/2}}
\newc{\mee}{m_{\beta\beta}}
\newc{\Ge}{^{76}\textrm{Ge}}
\newc{\bbbar}{B^0_d\textrm{-}\bar{B}^0_d}
\newc{\kkbar}{K^0\textrm{-}\bar{K^0}}
\newc{\solar}{\Delta m^2_{\odot}}
\newc{\atmos}{\Delta m^2_{atm}}
\newc{\mnu}{m_{\nu}}
\newc{\msusy}{\Lambda_{\textrm{SUSY}}}
\newc{\tick}{\surd}

\newc{\lam}{\lambda}

\newc{\ssup}{\tilde{\bar{U}}}
\newc{\ssdown}{\tilde{\bar{D}}}
\newc{\ssstrange}{\tilde{s}}
\newc{\sscharm}{\tilde{c}}
\newc{\sstop}{\tilde{t}}
\newc{\ssbottom}{\tilde{b}}

\newc{\sse}{\tilde{\bar{E}}}
\newc{\ssmu}{\tilde{\mu}}
\newc{\sstau}{\tilde{\tau}}
\newc{\ssnue}{\tilde{\nu_{e}}}
\newc{\ssnumu}{\tilde{\nu_{\mu}}}
\newc{\ssnutau}{\tilde{\nu_{\tau}}}
\newc{\ssbnue}{\bar{\tilde{\nu_{e}}}}
\newc{\ssbnumu}{\bar{\tilde{\nu_{\mu}}}}
\newc{\ssbnutau}{\bar{\tilde{\nu_{\tau}}}}
\newc{\ssnu}{\tilde{\nu}}

\newc{\sslep}{\tilde{L}}
\newc{\ssq}{\tilde{Q}}
\newc{\ssh}{\tilde{h}}

\newc{\mM}{\mathcal{M}}
\newc{\lag}{\mathcal{L}}
\newc{\sig}{\sigma}

\def\lsim{\raise0.3ex\hbox{$\;<$\kern-0.75em\raise-1.1ex\hbox{$\sim\;$}}}
\def\gsim{\raise0.3ex\hbox{$\;>$\kern-0.75em\raise-1.1ex\hbox{$\sim\;$}}}

\title{LHC and B physics probes of neutrinoless double beta decay in
  supersymmetry without R-parity} 
\author{B.~C.~Allanach$^1$, C.~H.~Kom$^{1,2}$, H.~P\"as$^3$
  \\$^1$ DAMTP, University of Cambridge, Wilberforce
  Road, Cambridge, CB3 0WA, UK
  \\$^2$ Cavendish Laboratory, J.J. Thomson Avenue, 
  Cambridge CB3 0HE, UK
  \\$^3$ Fakult\"at f\"ur Physik, Technische Universit\"at 
Dortmund, D-44221, Dortmund, Germany\\
{\rm E-mails:} \email{b.c.allanach@damtp.cam.ac.uk},
 \email{kom@hep.phy.cam.ac.uk},\\
{\phantom{E-mails: }}\email{heinrich.paes@uni-dortmund.de}
}
\preprint{CAVENDISH-HEP-2009-01\\DAMTP-2008-86\\DO-TH-08/10}

\abstract{In the event of an observation of neutrinoless double beta
  decay, a relevant question would be: what lepton number violating physics
  is responsible for the decay?
  The exchange of Majorana neutrinos and/or supersymmetric particles may
  contribute.
  We point out that measurements of supersymmetric signals at the LHC, including
  single slepton 
  production, could be used to help bound some supersymmetric processes
  contributing to neutrinoless double beta decay.
  LHC information about the supersymmetric spectrum could be combined with
  $\bbbar$ mixing data in order to bound a competing neutrinoless double beta
  decay process involving sbottom exchange.
}

\begin{document}

\tableofcontents

\section{Introduction}

Neutrinoless double beta decay ($\betadecay$) provides the most sensitive probe of
lepton number violation as well as a Majorana nature of neutrinos.
At the quark level, $\betadecay$ corresponds to the simultaneous transition
of two down quarks into two up-quarks and two electrons.
While the most prominent mechanism in the literature 
that causes this decay involves the exchange
of a massive Majorana neutrino, several other possibilities have been
discussed. 
Here, we focus on the attractive alternative where $\betadecay$ is
mediated by the exchange of sparticles in supersymmetric models
with R-parity violation
\cite{Hirsch:1995ek,Faessler:1998qv,Hirsch:1995cg,Pas:1998nn,Faessler:2007nz}. 

R-parity violating couplings arise in a
general supersymmetric extension of the Standard Model (SM), 
where the  
superpotential contains renormalisable baryon- and lepton- number violating operators.  The presence of both sets of operators typically leads 
to violation of stringent bounds on proton decay\cite{Goity:1994dq}, unless the parameters are 
unnaturally suppressed.  Proton decay bounds are evaded if a discrete symmetry 
is imposed which forbids at least one set of such parameters.  A 
widely-studied example is R-parity\cite{Farrar:1978xj}, under which both sets of 
operators are odd under the parity transformation and hence are forbidden.  
This symmetry also has the advantage of having the lightest supersymmetric 
particle (LSP) as a natural dark matter (DM) candidate.  On the other hand, 
there exist R-parity conserving (RPC) dimension 5 operators which could 
potentially lead to fast proton decay \cite{Sakai:1981pk,Weinberg:1981wj}.
One way to suppress these operators is by instead imposing proton hexality
\cite{Ibanez:1991pr,Dreiner:2005rd}. 

Instead of R-parity, we focus on an alternative, namely baryon 
triality (also known as baryon parity in some literature)
\cite{Ibanez:1991hv,Dreiner:2007uj}.  This $\mathbf{Z}_3$ discrete symmetry allows for the dimension 4 R-parity 
violating (RPV) terms which violate lepton number, while those that 
violate baryon number are forbidden.  A significant advantage of this 
class of models is that dimension 5 proton decay operators are forbidden.  
The LSP will decay via the non-zero RPV couplings present, hence a neutralino
LSP cannot be a dark matter candidate.  Other dark matter
candidates are viable, for example
the gravitino, since its decay is 
slow on cosmological time scales \cite{Buchmuller:2007ui}. Alternatively the
dark matter could originate from a hidden sector.  In the rest of this paper,
we use the 
term R-parity violation to denote 
the lepton number violating, R-parity violating interactions.
A survey of effective lepton number violating operators 
may be found for example in \cite{deGouvea:2007xp}.

After imposing a symmetry that forbids baryon-number violating terms, 
the RPV superpotential is
\eq 
\mathcal{W}_{RPV}=\frac{1}{2}\lam_{ijk}L_iL_j\bar{E}_k
+\lam'_{ijk}L_iQ_j\bar{D}_k -\mu_iL_iH_u, \label{superpot}
\qe
where we have suppressed all gauge indices and used the notation of
Ref.~\cite{Allanach:2001kg}. 
The $\{ i,j,k \} \in \{1,2,3\}$ are family indices.  
$\lambda_{ijk}$, $\lambda'_{ijk}$  are dimensionless trilinear
RPV couplings, and $\mu_i$ are bi-linear RPV parameters, having dimensions of
mass.   

Compared with the RPC minimal supersymmetric extension to the Standard Model
(RPC MSSM), the presence of additional RPV couplings leads to distinctive
signatures at a collider
and have
interesting physical consequences, for example providing neutrino masses. 
As the RPV couplings violate lepton number by 1 and
involve lepton doublets, they automatically lead to Majorana neutrino
masses
\cite{Joshipura:1994ib,Nowakowski:1995dx,Grossman:1997is,Grossman:1998py,Joshipura:1999hr,Davidson:2000uc,Davidson:2000ne,Grossman:2003gq,Dedes:2006ni}
without the need to introduce additional field content such as a
right handed neutrino. $\betadecay$ proceeds through the `1-1' entry in the
neutrino Majorana mass matrix $\mee$ 
 which connects two electron neutrinos in a basis 
where the charged lepton mass matrix is diagonal.
We refer to this mechanism as \emph{$\mee$ contribution} in the 
remainder of the paper.
For three left-handed Majorana neutrino masses $m_i$, with PMNS mixing matrix
$U_{\alpha i}$, where $\alpha \in \{ e, \mu, \tau\}$ and $i \in \{1,2,3\}$,
\eq
\mee = \sum_{i=1}^3 m_i U_{ei}^2.
\qe
$\mee$ may be a complex quantity. 
There are additional
higher dimensional effective operators, characteristic of this class of models,
which mediate  
$\betadecay$ without an $\mee$ insertion.  
As these operators violate lepton number by two units without the need of an 
$\mee$
insertion, they are not suppressed by the smallness of the neutrino masses as
a result.  These channels will be called \emph{direct contributions.}
In general, the direct and $m_{\beta \beta}$ contributions
contribute to $\betadecay$ simultaneously, so 
depending on their relative magnitudes, interference between the 
direct and $\mee$ decay matrix elements may need to be included in determining
the  decay rate of $\betadecay$.

A measurement of the $\betadecay$ rate alone 
does not fix a neutrino mass scale, since
it is possible that direct contributions are non-negligible.
However, the 
RPV couplings leading to the direct contributions can affect other
observables, which could then be used to constrain the
amplitude of 
the direct contributions. If they can be experimentally bounded, one
may infer $m_{\beta \beta}$ from the $\betadecay$ decay rate.

The aim of this paper is to explore the interplay between direct and
$m_{\beta \beta}$ contributions, in particular how different constraints and
observations may  
shed light on the underlying mechanisms of $\betadecay$.  
We will show how the combined knowledge of the masses of the
electron sneutrino and sbottoms and constraints from $\bbbar$ mixing
could allow us to determine an upper bound on a direct $\betadecay$ channel
involving sbottoms.  Searches for
single slepton production at the LHC could provide valuable
information on the value of $\lam'_{111}$, which (with measurements of
various sparticle masses) will allow 
one to bound its direct contribution to $\betadecay$.
A first exploration on the relationship between $\betadecay$ decay rate and
single slepton production at the LHC may be found in \cite{AKPprl}.  
A complementary way to probe mechanisms of $\betadecay$ was discussed in
\cite{Deppisch:2006hb,Gehman:2007qg}  
by combining measurements from different nuclei.  Efforts to relate neutrino
masses and collider phenomenology in other theories with lepton number
violation may be found for instance in \cite{AristizabalSierra:2007nf}. 

For concreteness, we only consider $\betadecay$ of $\Ge$.  In the rest of this
paper, numerical values of nuclear matrix elements (NMEs) and half life all
refer to this nucleus.  
A related work on contributions of trilinear RPV terms to
$\betadecay$ 
is presented in \cite{Wodecki:1999gu}.
There, particular attention is paid to nuclear matrix element
calculations and contributions from different sparticles in the presence of a
non-zero $\lam'_{111}$ coupling.  
We go beyond the scope of this work in several ways: most importantly, 
we focus on what may be inferred from different experimental measurements
on the mechanism that produces $\betadecay$. 
We have also corrected certain terms in
the effective Lagrangian at the quark level, that were incorrect in
the literature.

This paper is organised as follows.  
In section \ref{sec:0vbbSummary} we briefly review the possible
$\betadecay$ mechanisms in the RPV MSSM.   
The current experimental half life limit $\bhalflife$ of $\Ge$, as well as the 
neutrino oscillation data are summarised in section \ref{sec:exptLimit}.  
We also list some useful scaling relations between the RPV parameter bounds
and SUSY breaking parameters  used there.  In section
\ref{sec:lamp113131} we proceed to discuss how the RPV contribution to
$\bbbar$ mixing can affect the possibility of the
$\lam'_{113}\lam'_{131}$ direct contribution to be the dominant observable 
$\betadecay$ channel.  We then investigate the prospects of observing 
the single selectron resonance at the LHC and its implication 
of $\betadecay$ in 
section \ref{sec:lamp111}, before concluding in section \ref{sec:disc}.
Technical information about parton-level $\betadecay$ calculations is in
Appendix~\ref{sec:part} and
our NMEs are listed
in Appendix~\ref{nmes}.

\section{Mechanisms of $\betadecay$ in the RPV MSSM} \label{sec:0vbbSummary}
In the RPV MSSM, it is possible to construct Majorana
neutrino mass models that 
explain the neutrino oscillation data
\cite{Dedes:2006ni,Dreiner:2007uj,Rakshit:1998kd,Joshipura:1999hr,Abada:2000xr,Abada:2002ju,Hirsch:2000ef,Diaz:2003as,Allanach:2007qc}.
$\betadecay$ could proceed through 
standard 
light Majorana neutrino exchange with the $\mee$ mass insertion.  
In addition, there are direct contributions via 
the RPV coupling products 
$\lam'_{113}\lam'_{131}$ and $\lam'_{111}\lam'_{111}$. 

\begin{figure}[!ht]
  \begin{center}
    {
      \begin{picture}(210,75)(0,0)
	{
	  \DashArrowArc(105,40)(20,90,180){3}
	  \DashArrowArcn(105,40)(20,90,0){3}
	  \ArrowArcn(105,40)(20,270,180)
	  \ArrowArc(105,40)(20,270,360)
	  \ArrowLine(45,40)(85,40)
	  \ArrowLine(165,40)(125,40)
	  \Vertex(105,20){1.5}
	  \Vertex(105,60){1.5}
	  \put(35,39){$\nu_e$}
	  \put(167,39){$\nu_e$}
	  \put(60,28){$\lambda'_{113}$}
	  \put(128,28){$\lambda'_{131}$}
	  \put(98,65){$(m^{2*}_{\tilde{b}})_{LR}$}
	  \put(100,10){$m^*_{d}$}
	}
      \end{picture}
    }
    \caption[1 loop correction to $\mee$ in mass insertion approximation]{An
      example diagram showing a contribution to $\mee$ from the product
      $\lam'_{113}\lam'_{131}$.} 
    \label{fig:mee}
  \end{center}
\end{figure}
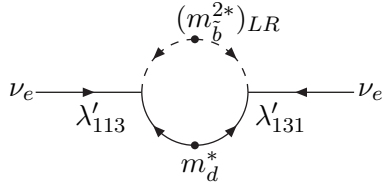

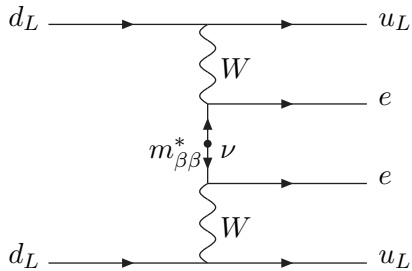
\begin{figure}[!ht] \begin{center}
    \begin{picture}(180,150)(0,0)
      \ArrowLine(30,30)(90,30)
      \ArrowLine(90,30)(150,30)
      \ArrowLine(30,120)(90,120)
      \ArrowLine(90,120)(150,120)
      \ArrowLine(90,60)(150,60)
      \ArrowLine(90,90)(150,90)
      \ArrowLine(90,75)(90,60)
      \ArrowLine(90,75)(90,90)
      \Photon(90,30)(90,60){3}{2.5}
      \Photon(90,90)(90,120){3}{2.5}
      \Vertex(90,75){1.5}
      \put(15,30){$d_L$}
      \put(15,120){$d_L$}
      \put(155,120){$u_L$}
      \put(155,90){$e$}
      \put(155,60){$e$}
      \put(155,30){$u_L$}
      \put(95,70){$\nu$}
      \put(68,70){$\mee^*$}
      \put(95,100){$W$}
      \put(95,40){$W$}
    \end{picture}
  \caption{A Feynman diagram showing $\betadecay$ via exchange of a virtual
    Majorana neutrino.}
  \label{fig:mee0vbb}
\end{center}
\end{figure}

The RPV coupling products $\lam'_{ilm}\lam'_{jml}$ contribute to the neutrino
mass matrix $(\mnu)_{ij}$, in particular they generate 
the `1-1' entry $\mee$.  A Feynman diagram using the mass insertion
approximation (MIA) \cite{Misiak:1997ei} with $\lam'_{113}\lam'_{131}$ is
shown in fig.~\ref{fig:mee}. 
In this approximation, $\mee$ can be written as
\eq\label{eq:mee}
\mee \simeq  \frac{3m_d}{8\pi^2}\frac{\lambda'_{113}\lambda'_{131}m^2_{\tilde{b}_{LR}}}{m^2_{\tilde{b}_{LL}}-m^2_{\tilde{b}_{RR}}}\textrm{ln}\Big(\frac{m^2_{\tilde{b}_{LL}}}{m^2_{\tilde{b}_{RR}}}\Big) 
+ (b \leftrightarrow d).
\qe
Here $m^{2}_{\tilde{b}_{LL}}$, $m^{2}_{\tilde{b}_{RR}}$ and
$m^{2}_{\tilde{b}_{LR}}$ represent the entries of the sbottom mass matrix in
an obvious notation, and $m_d$ is the running mass of the down quark.
$\betadecay$ may then proceed via exchange of a virtual Majorana neutrino
with a $\mee$ insertion.  
A Feynman diagram depicting this process is displayed in fig.~\ref{fig:mee0vbb}.
For a realistic model, 
one expects there to be many non-vanishing bi-linear and/or tri-linear RPV
operators 
present in order to 
fill out the effective 3 $\times$ 3 neutrino mass matrix with non-zero
entries.  Bi-linear RPV couplings could also lead to direct $\betadecay$.  However, as discussed in \cite{Hirsch:2000jt}, neutrino mass terms obtained from these couplings that are consistent with the observed neutrino mass scales typically lead to negligible direct contributions to the $\betadecay$ rate.  On this basis we neglect their direct contribution, but bear in mind the possibility that they may enhance $\mee$ beyond what is expected from the tri-linear couplings.  We also expect that the $\lam_{ijk}$ couplings may contribute to $\mee$, but not to affect direct $\betadecay$ significantly.  Coupling products of the form $\lam'_{11k}\lam'_{ik1}$ and $\lam'_{k11}\lam_{i1k}$ could violate lepton number $i$ and electron number by 1 unit each.  For $i \ne 1$, their contributions to direct $\betadecay$ via PMNS mixing are suppressed by the mass scale of the light neutrinos.  As a result their contributions to direct $\betadecay$ is likely to be subdominant and will not be discussed further in this paper.

A Feynman diagram representing a direct contribution mediated by
$\lam'_{113}\lam'_{131}$ 
is shown in fig.~\ref{fig:lamp1130vbb}.  As the corresponding matrix 
element does not contain neutrino mass insertions, it is not suppressed by the
smallness of the neutrino mass scale.
\begin{figure}[!ht] \begin{center}
    \begin{picture}(180,150)(0,0)
      \ArrowLine(30,30)(60,30)
      \ArrowLine(60,30)(150,30)
      \ArrowLine(60,60)(150,60)
      \ArrowLine(60,105)(60,60)
      \ArrowLine(60,105)(30,105)
      \ArrowLine(120,105)(150,120)
      \ArrowLine(120,105)(150,90)
      \DashArrowLine(60,105)(90,105){3.5}
      \DashArrowLine(120,105)(90,105){3.5}
      \Photon(60,30)(60,60){3}{3.5}
      \Vertex(90,105){1.5}
      \put(15,30){$d_L$}
      \put(15,105){$d^c$}
      \put(155,30){$u_L$}
      \put(155,60){$e_L$}
      \put(155,90){$e_L$}
      \put(155,120){$u_L$}
      \put(40,45){$W_{\mu}$}
      \put(45,82.5){$\nu_e$}
      \put(75,90){$\tilde{b}_{LL}$}
      \put(105,90){$\tilde{b}_{RR}$}
      \put(55,115){$\lam'^*_{131}$}
      \put(78,115){$m^2_{\tilde{b}_{LR}}$}
      \put(112,115){$\lam'^*_{113}$}
    \end{picture}
  \caption{Direct $\lam'_{113}\lam'_{131}$ contribution to $\betadecay$.}
  \label{fig:lamp1130vbb}
\end{center}
\end{figure}
In principle, the 
product $\lam'_{112}\lam'_{121}$ could also lead to $\betadecay$.  However, 
this coupling product is tightly constrained by $\kkbar$ mixing 
\cite{Choudhury:1996ia}, and hence this contribution is 
neglected in the rest of this paper.  

It should be noted that $\lam'_{111}\lam'_{111}$ can also mediate
$\betadecay$ via a diagram similar to fig.~\ref{fig:lamp1130vbb}.
However, in the case of $\lam'_{111}\lam'_{111}$, there are other
diagrams contributing to $\betadecay$ which dominate. These diagrams are shown
in fig.~\ref{fig:lamp1110vbb}.

\begin{figure}[!ht]
  \begin{center}
    \scalebox{0.73}{
      \begin{tabular}{ccc}
	\subfigure[]{
	  \begin{picture}(180,150)(0,0)
	    \ArrowLine(60,45)(30,45)
	    \ArrowLine(60,45)(60,75)
	    \ArrowLine(60,105)(60,75)
	    \ArrowLine(60,105)(30,105)
	    \ArrowLine(120,45)(150,30)
	    \ArrowLine(120,45)(150,60)
	    \ArrowLine(120,105)(150,90)
	    \ArrowLine(120,105)(150,120)
	    \DashLine(120,45)(60,45){3.5}
	    \DashLine(120,105)(60,105){3.5}
	    \Vertex(60,75){1.5}
	    \put(15,45){$d^c$}
	    \put(15,105){$d^c$}
	    \put(155,30){$u_L$}
	    \put(155,60){$e_L$}
	    \put(155,90){$e_L$}
	    \put(155,120){$u_L$}
	    \put(35,75){$m_{\chi,\tilde{g}}$}
	    \put(90,30){$\tilde{d_R}$}
	    \put(90,115){$\tilde{d_R}$}
	    \put(110,52){$\lam'^*_{111}$}
	    \put(110,93){$\lam'^*_{111}$}
	  \end{picture}\label{fig:lamp111a}
	}
	&    
	\subfigure[]{
	  \begin{picture}(180,150)(0,0)
	    \ArrowLine(90,30)(30,30)
	    \ArrowLine(90,30)(150,30)
	    \ArrowLine(90,120)(30,120)
	    \ArrowLine(90,120)(150,120)
	    \ArrowLine(90,60)(150,60)
	    \ArrowLine(90,90)(150,90)
	    \ArrowLine(90,60)(90,75)
	    \ArrowLine(90,90)(90,75)
	    \DashLine(90,30)(90,60){3.5}
	    \DashLine(90,90)(90,120){3.5}
	    \Vertex(90,75){1.5}
	    \put(15,30){$d^c$}
	    \put(15,120){$d^c$}
	    \put(155,120){$e_L$}
	    \put(155,90){$u_L$}
	    \put(155,60){$u_L$}
	    \put(155,30){$e_L$}
	    \put(95,70){$\tilde{\chi}/\tilde{g}$}
	    \put(65,70){$m_{\chi/\tilde{g}}$}
	    \put(95,100){$\tilde{u_L}$}
	    \put(95,40){$\tilde{u_L}$}
	    \put(85,125){$\lam'^*_{111}$}
	    \put(85,20){$\lam'^*_{111}$}
	  \end{picture}\label{fig:lamp111b}
	}
        &
	\subfigure[]{
	  \begin{picture}(180,150)(0,0)
	    \ArrowLine(60,45)(30,45)
	    \ArrowLine(60,45)(60,67.5)
	    \ArrowLine(60,90)(60,67.5)
	    \ArrowLine(60,90)(150,90)
	    \ArrowLine(60,120)(150,120)
	    \ArrowLine(60,120)(30,120)
	    \ArrowLine(120,45)(150,30)
	    \ArrowLine(120,45)(150,60)
	    \DashLine(60,45)(120,45){3.5}
	    \DashLine(60,90)(60,120){3.5}
	    \Vertex(60,67.5){1.5}
	    \put(15,45){$d^c$}
	    \put(15,120){$d^c$}
	    \put(155,120){$e_L$}
	    \put(155,30){$u_L$}
	    \put(155,60){$e_L$}
	    \put(155,90){$u_L$}
	    \put(30,67.5){$m_{\chi/\tilde{g}}$}
	    \put(65,67.5){$\tilde{\chi}/\tilde{g}$}
	    \put(90,30){$\tilde{d_R}$}
	    \put(65,105){$\tilde{u_L}$}
	    \put(55,125){$\lam'^*_{111}$}
	    \put(110,52){$\lam'^*_{111}$}
	  \end{picture}\label{fig:lamp111c}
	}
	\\
	\subfigure[]{
	  \begin{picture}(180,150)(0,0)
	    \ArrowLine(90,30)(30,30)
	    \ArrowLine(90,30)(150,30)
	    \ArrowLine(90,120)(30,120)
	    \ArrowLine(90,120)(150,120)
	    \ArrowLine(90,60)(150,60)
	    \ArrowLine(90,90)(150,90)
	    \ArrowLine(90,60)(90,75)
	    \ArrowLine(90,90)(90,75)
	    \DashLine(90,30)(90,60){3.5}
	    \DashLine(90,90)(90,120){3.5}
	    \Vertex(90,75){1.5}
	    \put(15,30){$d^c$}
	    \put(15,120){$d^c$}
	    \put(155,120){$u_L$}
	    \put(155,90){$e_L$}
	    \put(155,60){$e_L$}
	    \put(155,30){$u_L$}
	    \put(95,70){$\tilde{\chi}$}
	    \put(70,70){$m_{\chi}$}
	    \put(95,100){$\tilde{e_L}$}
	    \put(95,40){$\tilde{e_L}$}
	    \put(85,125){$\lam'^*_{111}$}
	    \put(85,20){$\lam'^*_{111}$}
	  \end{picture}\label{fig:lamp111d}
	}
        &
	\subfigure[]{
	  \begin{picture}(180,150)(0,0)
	    \ArrowLine(60,45)(30,45)
	    \ArrowLine(60,45)(60,67.5)
	    \ArrowLine(60,90)(60,67.5)
	    \ArrowLine(60,90)(150,90)
	    \ArrowLine(60,120)(150,120)
	    \ArrowLine(60,120)(30,120)
	    \ArrowLine(120,45)(150,30)
	    \ArrowLine(120,45)(150,60)
	    \DashLine(60,45)(120,45){3.5}
	    \DashLine(60,90)(60,120){3.5}
	    \Vertex(60,67.5){1.5}
	    \put(15,45){$d^c$}
	    \put(15,120){$d^c$}
	    \put(155,120){$u_L$}
	    \put(155,30){$u_L$}
	    \put(155,60){$e_L$}
	    \put(155,90){$e_L$}
	    \put(40,67.5){$m_{\chi}$}
	    \put(65,67.5){$\tilde{\chi}$}
	    \put(90,30){$\tilde{d_R}$}
	    \put(65,105){$\tilde{e_L}$}
	    \put(55,125){$\lam'^*_{111}$}
	    \put(110,52){$\lam'^*_{111}$}
	  \end{picture}\label{fig:lamp111e}
	}
	&
	\subfigure[]{
	  \begin{picture}(180,150)(0,0)
	    \ArrowLine(90,30)(30,30)
	    \ArrowLine(90,30)(150,30)
	    \ArrowLine(90,120)(30,120)
	    \ArrowLine(90,120)(150,120)
	    \ArrowLine(90,60)(150,60)
	    \ArrowLine(90,90)(150,90)
	    \ArrowLine(90,60)(90,75)
	    \ArrowLine(90,90)(90,75)
	    \DashLine(90,30)(90,60){3.5}
	    \DashLine(90,90)(90,120){3.5}
	    \Vertex(90,75){1.5}
	    \put(15,30){$d^c$}
	    \put(15,120){$d^c$}
	    \put(155,120){$u_L$}
	    \put(155,90){$e_L$}
	    \put(155,60){$u_L$}
	    \put(155,30){$e_L$}
	    \put(95,70){$\tilde{\chi}$}
	    \put(70,70){$m_{\chi}$}
	    \put(95,100){$\tilde{e_L}$}
	    \put(95,40){$\tilde{u_L}$}
	    \put(85,125){$\lam'^*_{111}$}
	    \put(85,20){$\lam'^*_{111}$}
	  \end{picture}\label{fig:lamp111f}
	}
      \end{tabular}
    }
    \caption[$\lam'_{111}\lam'_{111}$ contributions to $\betadecay$]{$\lam'_{111}\lam'_{111}$ contributions to $\betadecay$.}\label{fig:lamp1110vbb}
  \end{center}
\end{figure}

\section{Experimental limits}\label{sec:exptLimit}
\begin{figure}[h!]
  \centering
  \includegraphics[width=0.5\textwidth]{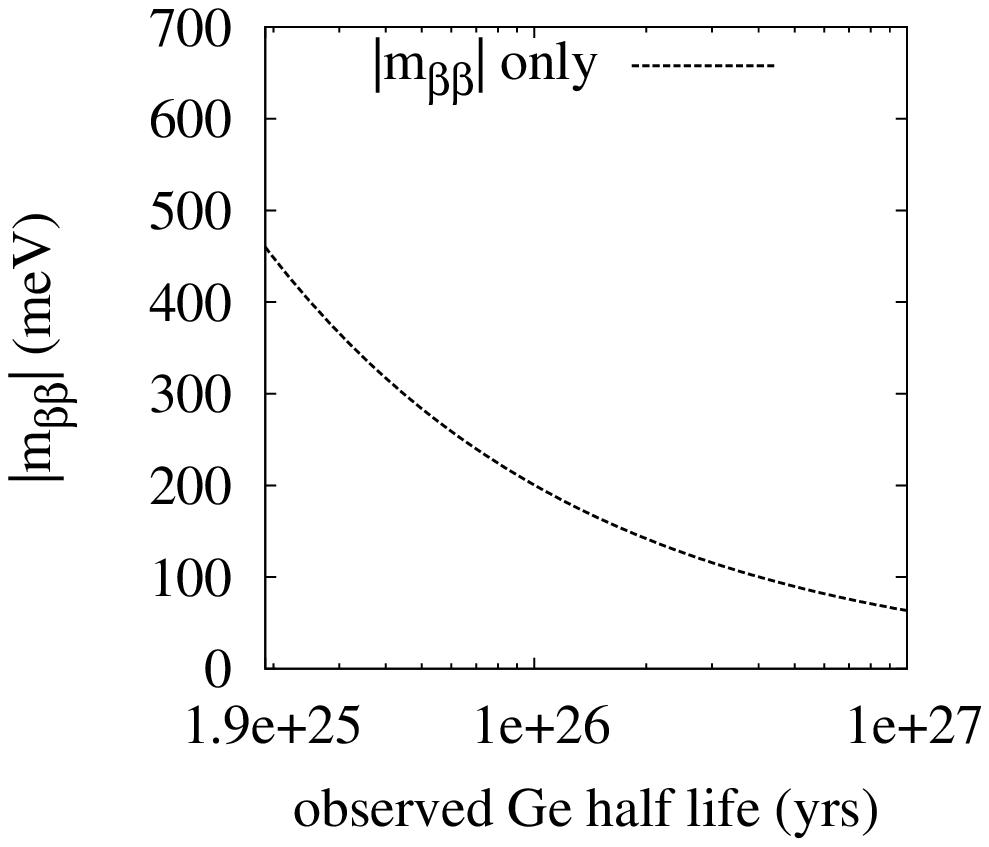}
  \caption{Value of $\mee$ inferred from a future observation of $\bhalflife$,
  for our NMEs and assuming only Majorana
  neutrino exchange contributes.}
\label{fig:meein}
\end{figure}

The most stringent lower limit on the $\Ge$ $\betadecay$ half life was bounded
by the Heidelberg-Moscow experiment \cite{Baudis:1999xd,KlapdorKleingrothaus:2000sn} to be
\eqa \label{eq:halflifelimit}
\bhalflife &\geq& 1.9\cdot 10^{25} \textrm{yrs}.
\qea
The most common interpretation of such a limit is in terms of a model in which
only 
$\mee$ contributes to $\betadecay$. 
In the event of an observation in the next round of $\betadecay$ searches, the
half-life will be used to infer $|\mee|$, as in fig.~\ref{fig:meein}. The figure
shows that in the event of an observation of $\bhalflife<10^{27}$ years,
$|\mee|>50$ meV. $|\mee| \gsim 10$ meV implies a non-hierarchical (i.e.\ inverted or quasi-degenerate)
spectrum~\cite{Amsler:2008zzb}.

Such inferences depend crucially on the assumption that no other process (for
example, direct RPV processes) contribute to $\betadecay$. We shall now take
into account the possibility that the direct RPV processes can simultaneously
contribute to $\mee$. 
Depending on the RPV couplings considered, different mechanisms
dominate the $\betadecay$ process.  Thus eq.~(\ref{eq:halflifelimit}) can be translated to upper bounds on particular products of RPV couplings.  For the model under consideration, the relevant formula is given by 
\eq \label{eq:halflifeformula}
(\bhalflife)^{-1}  =G_{01} \left|M_{tot}\right|^2=G_{01}\left|
|\frac{\mee}{m_e}M_{\nu}|
+e^{i \phi_1}| M_{\lam'_{113}\lam'_{131}}|
  + e^{i \phi_2} |M_{\lam'_{111}}| \right|^2, 
\qe
where $G_{01}=7.93\phantom{0}10^{-15}\textrm{yr}^{-1}$
\cite{Pantis:1996py} is a precisely calculable phase space factor and $e^{i\phi_{1,2}}$
are relative complex phases between the various contributions.  These
matrix elements represent contributions from the direct
($M_{\lam'_{111}},M_{\lam'_{113}\lam'_{131}}$) and the neutrino mass
($M_{\nu}$) mechanisms.  

If $\mee$ is the dominant contribution to $\betadecay$, 
using the value of $M_\nu$ displayed in Appendix~\ref{nmes},
the bound in eq.~(\ref{eq:halflifelimit}) can be translated into the limit
\eqa \label{eq:meebound}
\mee &\lesssim& 460 \textrm{ meV}.
\qea
If instead the direct contributions are dominant, then assuming the $\mee$
contributions are negligible, eq.~(\ref{eq:halflifelimit}) leads to
\cite{Pas:1998nn,Faessler:2007nz,Hirsch:1995ek,Faessler:1998qv} 
\eqa \label{eq:113directdecaylimit}
\lam'_{113}\lam'_{131} &\lesssim& 2\cdot 10^{-8} \Big(\frac{\Lambda_{SUSY}}{100\textrm{GeV}}\Big)^{3}, \\
\lam'_{111} &\lesssim& 5 \cdot 10^{-4}\Big(\frac{m_{\tilde{f}}}{100\textrm{GeV}}\Big)^{2}\Big(\frac{m_{\tilde{g}/\tilde{\chi}}}{100\textrm{GeV}}\Big)^{1/2},\label{eq:111directdecaylimit}
\qea
respectively, after taking into account the modifications to the effective
Lagrangians discussed in Appendix~\ref{sec:part}.  Here $\Lambda_{SUSY}$ is an
effective SUSY breaking scale for the soft terms involved in
eq.~(\ref{eq:eta}), and $m_{\tilde{f}}$ and $m_{\tilde{g}/\tilde{\chi}}$ are
sfermions and gluino/neutralino masses in the dominant Feynman diagrams. 

Clearly, the bounds on the RPV couplings depend on the SUSY mass spectra.  We
thus make a simple assumption about the RPC soft SUSY breaking terms: 
that they follow the 
minimal supergravity (mSUGRA) boundary
conditions.  The resulting SUSY mass spectra are obtained using the spectrum generator \softsusy\cite{Allanach:2001kg}.  
At the SUSY scale, all RPV couplings are set to zero except for either
$\lam'_{111}$ or $\lam'_{113}$ and $\lam'_{131}$.  
This allows
us to emphasise effects from the above RPV couplings, without additional
complications.
Specifically, the following set of parameters are defined: 
\eq\label{eq:0vbbmSUGRAspace}
\MO=[40,1000]\mbox{~GeV}, \
\Mhalf=[40,1000]\mbox{~GeV},\ 
\AO=0, \
\tanb=10,\ \sgnmu=+1,
\qe
where $\MO$, $\Mhalf$ and $\AO$ are the universal scalar, gaugino, and
trilinear soft SUSY breaking parameters defined at a unification scale
$M_{X}\sim 2.0\cdot 10^{16}\textrm{GeV}$, $\tanb$ is the ratio of the Higgs
vacuum expectation values $v_u/v_d$, and $\sgnmu$ is the sign of the bi-linear
Higgs parameter in the superpotential.

\begin{figure}[!ht]
  \centering
  \includegraphics[width=0.65\textwidth]{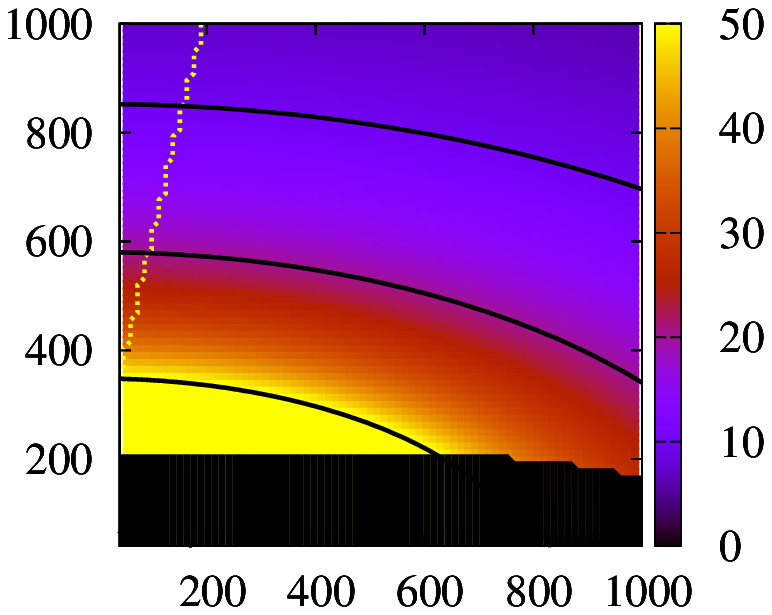}
  \put(-155,10){$\MO$/GeV}
  \put(-240,80){\rotatebox{90}{$\Mhalf$/GeV}}
  \put(-80,167){$\mathcal{R}$}
  \put(-150,75){\small{$\mathcal{R}=50$}}
  \put(-130,103){\White{\small{$\mathcal{R}=20$}}}
  \put(-117,137){\White{\small{$\mathcal{R}=10$}}}
  \label{fig:113R_NMERPVmee_Faessler.eps}
  \caption[Comparison of direct RPV and RPV-induced $\mee$ contributions to $\betadecay$ in a mSUGRA plane]{
    Ratio $\mathcal{R}=|M_{\lam'_{113}\lam'_{131}}/M_{\mee}|$, where
    $M_{\mee}=(\mee/m_e)M_{\nu}$, in RPV contributions to neutrinoless double
    beta decay.  
The blacked region out at the bottom of the plot is excluded and the dotted
line delimits regions of different LSP (see text).} \label{fig:NMERPVmeeRatioNew}
\end{figure}

We now turn to interference between
different contributions to the $\betadecay$ rate and discuss first the case 
where both $\mee$ and a direct RPV contribution are due to the same product 
of RPV couplings.  In previous studies,
such interference was neglected.  For direct contributions, this is a good
approximation for a SUSY mass scale $\msusy$ of the order 100 GeV, as the
$\mee$ generated from the same RPV couplings is sub-dominant.  However, for
fixed RPV couplings, $M_{\lam'_{111}}$ scales as $\msusy^{-5}$,
$M_{\lam'_{113}\lam'_{131}}$ scales as $\msusy^{-3}$, whereas $\mee$ scales as
$\msusy^{-1}$.  Thus the $\mee$ contribution would dominate at high SUSY
breaking scales.  The ratio
$\mathcal{R}=|M_{\lam'_{113}\lam'_{131}}/M_{\mee}|$, where
$M_{\mee}=(\mee/m_e)M_{\nu}$ is shown in fig.~\ref{fig:NMERPVmeeRatioNew}.
The region in black at 
low $\Mhalf$ has either no electroweak symmetry breaking (EWSB), and/or the
lightest Higgs mass is in violation of the LEP2 direct search limits
\cite{Schael:2006cr}. 
The 
LEP2 95\% confidence level upper bound implies $m_h>114.4$ GeV, but we impose 
$m_h>111.4$ GeV in order to include a 3 GeV theory uncertainty in the 
\softsusy~prediction of $m_h$.  The yellow dotted line on the left
separates the region with a stau (left) and a neutralino (right) LSP.  We see
that while in the lower mass region the direct contributions dominate over the
$\mee$ contributions, they become comparable in the high mass region, where
interference cannot be neglected.  We thus include interference terms in the
calculations which follow. 

On the other hand, despite the $\msusy^{-5}$ dependence on $M_{\lam'_{111}}$,
the ratio $|M_{\lam'_{111}}/M_{\mee}|$ is greater than 20 
in our parameter space region.  This is because $\mee$
generated by $\lam'_{111}$ is heavily suppressed by the running up quark mass
insertions in the loop diagrams.  
For this reason, we may safely neglect $\mee$
contributions to $\betadecay$ when considering cases with non-zero
$\lam'_{111}$ unless it originates from some other coupling.

\section{$\bbbar$ mixing, single slepton production at the LHC and
  $\betadecay$ \label{sec:inference} }

\subsection{Implications of $\lam'_{113}\lam'_{131}$ bound from $\bbbar$ on $\betadecay$}\label{sec:lamp113131}

\begin{figure}[!ht]
  \centering
  \scalebox{1.0}{
    \includegraphics[width=0.45\textwidth, height=0.40\textwidth]{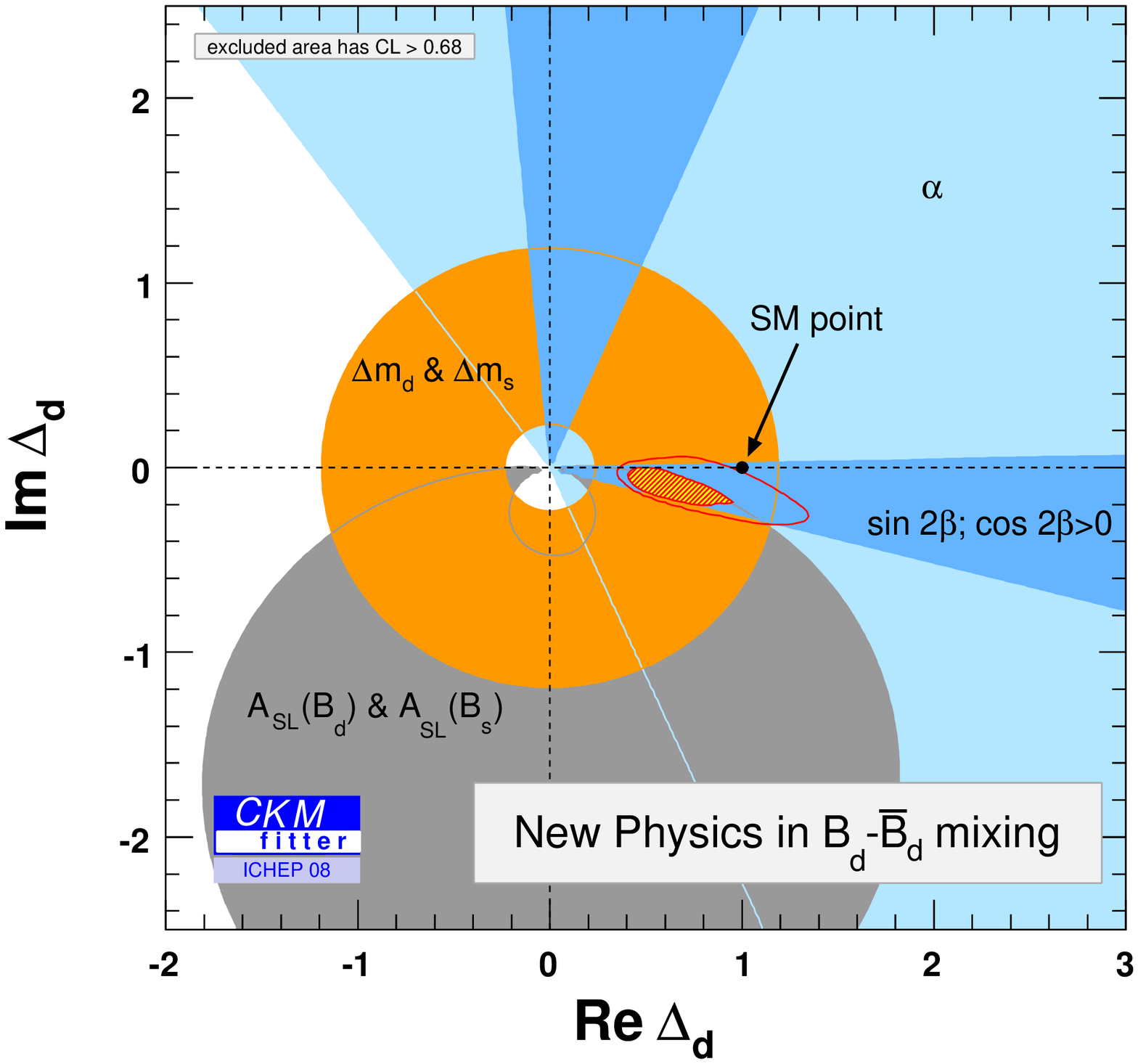}
    \label{fig:NPbbbarFit08a.eps}
  }  
  \caption[Possible new physics contributions to $\bbbar$.]{Possible new
    physics contributions to $\bbbar$, from Ref.~\cite{Charles:2008slides}.
    The 95\% C.L. region is shown in solid orange line.  The SM solution is
    located at $\Delta_d=1$, and deviation from the SM value may be attributed
    to the RPV contributions proportional to
    $\lam'_{113}\lam'_{131}$.}\label{fig:NPfits} 
\end{figure}

It was shown in \cite{Pas:1998nn} that for sparticle masses of $\sim$ 100 GeV,
a stringent bound on $\lam'_{113}\lam'_{131}$ comes from $\betadecay$.
Another competing bound comes from $\bbbar$ mixing.  A Feynman diagram of the
latter process is displayed in fig.~\ref{fig:bbbar}. 
\begin{figure}[!ht]
  \begin{center}
    {
      \begin{picture}(210,75)(0,0)
	{
	  \DashArrowLine(130,40)(80,40){3}
	  \ArrowLine(45,15)(80,40)
	  \ArrowLine(45,65)(80,40)
	  \ArrowLine(130,40)(165,15)
	  \ArrowLine(130,40)(165,65)
	  \put(100,30){$\tilde{\nu}_e$}
	  \put(35,15){$d$}
	  \put(35,65){$b^c$}
	  \put(170,15){$d^c$}
	  \put(170,65){$b$}
	  \put(80,45){$\lam'_{113}$}
	  \put(115,45){$\lam'^*_{131}$}
	}
      \end{picture}
    }
    \caption[$\bbbar$ mixing through sneutrino exchange via $\lam'_{113}\lam'_{131}$]{$\bbbar$ mixing through coupling product $\lam'_{113}\lam'_{131}$.}
    \label{fig:bbbar}
  \end{center}
\end{figure}
A recent update \cite{Deschamps:2008de} by the CKMfitter group shows that at 95\% confidence level, the magnitude of any new physics effect to $\bbbar$ mixing must be less than the SM contribution.  This result is shown in fig.~\ref{fig:NPfits}.  
In this figure, the Standard Model solution is located at $\Delta_d=1$, and
deviation from unity represents new physics effects contributing to $\bbbar$
mixing.   
The upper limit of
$\lam'_{113}\lam'_{131}$ is obtained by updating the results in
\cite{Pas:1998nn} to take into account the latest $\bbbar$ mixing data.
We obtain
\eqa\label{eq:bbbar}
\lam'_{113}\lam'_{131}\leq 4.0\cdot
10^{-8}\frac{m^2_{\tilde{\nu_e}}}{(100\textrm{GeV})^2}. 
\qea

To compare limits on $\lam'_{113}\lam'_{131}$ from $\bbbar$ and from $\betadecay$, we recall that the bound from $\betadecay$ depends on the sbottom mass squared matrix.  In the case where all SUSY breaking parameters are of the same order of magnitude, this bound relaxes approximately as the cube of the sbottom mass scale,
which is more rapid compared with the $\bbbar$ bound in eq.~(\ref{eq:bbbar}).
However in the most general MSSM the mass parameters relevant for these two
bounds are independent.  Which of the bounds is more stringent depends
therefore on the ratio of sneutrino to sbottom masses. 

In the following, we restrict our discussion to the parameter space discussed in
section \ref{sec:exptLimit}.  From \cite{Pas:1998nn}, we expect the
$\betadecay$ bound to be more stringent only in very low $\msusy$ of around
100 GeV.  With the new $\bbbar$ limit, and the fact that mSUGRA-like mass
spectra generally have squarks much heavier than the sleptons, we find that
the bound from $\bbbar$ mixing is more stringent than that from $\betadecay$
in all allowed regions of parameter space we explore.  This means that if the
direct $\lam'_{113}\lam'_{131}$ contribution is the only source of
$\betadecay$, then $\bhalflife$ must be larger than the current experimental
limit. This conclusion is sensitive to theoretical uncertainties in our
predicted value of $\bhalflife$ coming from the NMEs. 
Estimates in such uncertainties vary: NME calculations based on different
nuclear model assumptions and input parameters could differ by a factor of
1.3, 3 or even up to 5~\cite{Elliott:2004hr,Rodin:2007fz}. These are then
squared  
in order to obtain the half-life prediction. We take for example a factor of
3 uncertainty in the predicted half life, equivalent to $\pm 0.5$ in
$\log_{10} \bhalflife$. 
If
the NMEs predicted $\bhalflife$ 
to be a factor of 3 less than the values taken here, then there is a small
region 
where the stronger bound would instead originate from $\bhalflife$ at
the lowest viable values of $\Mhalf$ and $\MO/\mbox{GeV} \sim 400-600$.
See the appendix for details about the NMEs. 

\begin{figure}[h!]
  \centering
  \includegraphics[width=0.6\textwidth]{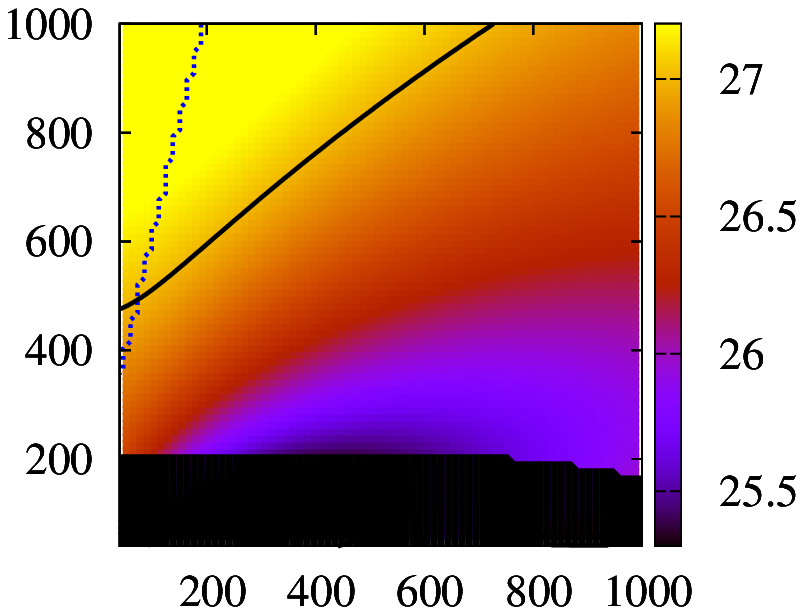}
  \put(-145,10){\small $M_0$/GeV}
  \put(-220,70){\rotatebox{90}{\small $M_{1/2}$/GeV}}
  \put(-90,157){\small $\textrm{log}_{10}\bhalflife$}
  \put(-160,100){\small $1.0 \cdot 10^{27}$ yrs}
  \caption[]{Lower limit on $\bhalflife(\Ge)$, using upper limit on
    $\lam'_{113}\lam'_{131}$ obtained from $\bbbar$ mixing.  The black contour
    shows where a $\bhalflife$ limit of $10^{27}$ years is expected. 
The black region at the bottom of the plot is excluded and the dotted
line delimits regions of different LSP (see text).}
\label{fig:lamp113bd}
\end{figure}


The variation of the $\bhalflife$ lower limit, obtained from an upper bound on
$\lam'_{113}\lam'_{131}$ from $\bbbar$ mixing is shown in
fig.~\ref{fig:lamp113bd}.  As in fig.~\ref{fig:NMERPVmeeRatioNew}, the (blue)
dotted line separates regions with stau LSP and neutralino LSP, while the
black region at the bottom is excluded due to no EWSB or the higgs being too
light.  We see that $\bhalflife \sim 10^{26}$--$10^{27}\textrm{yrs}$ is still 
allowed in much of the parameter space, so that $\betadecay$ can be detected 
by next generation experiments.  In particular, there exist good prospects 
of observing a $\betadecay$ signal in
the region with relatively low $\Mhalf$.  This is because the sbottom masses
receive large renormalisation effects from the gluino mass and as a result are
much lighter in the low $\Mhalf$ region.  This in turn enhances
$M_{\lam'_{113}\lam'_{131}}$ compared with corresponding values in the high
$\Mhalf$ region.


\begin{figure}[!ht]{
\begin{center}
\includegraphics[width=7cm]{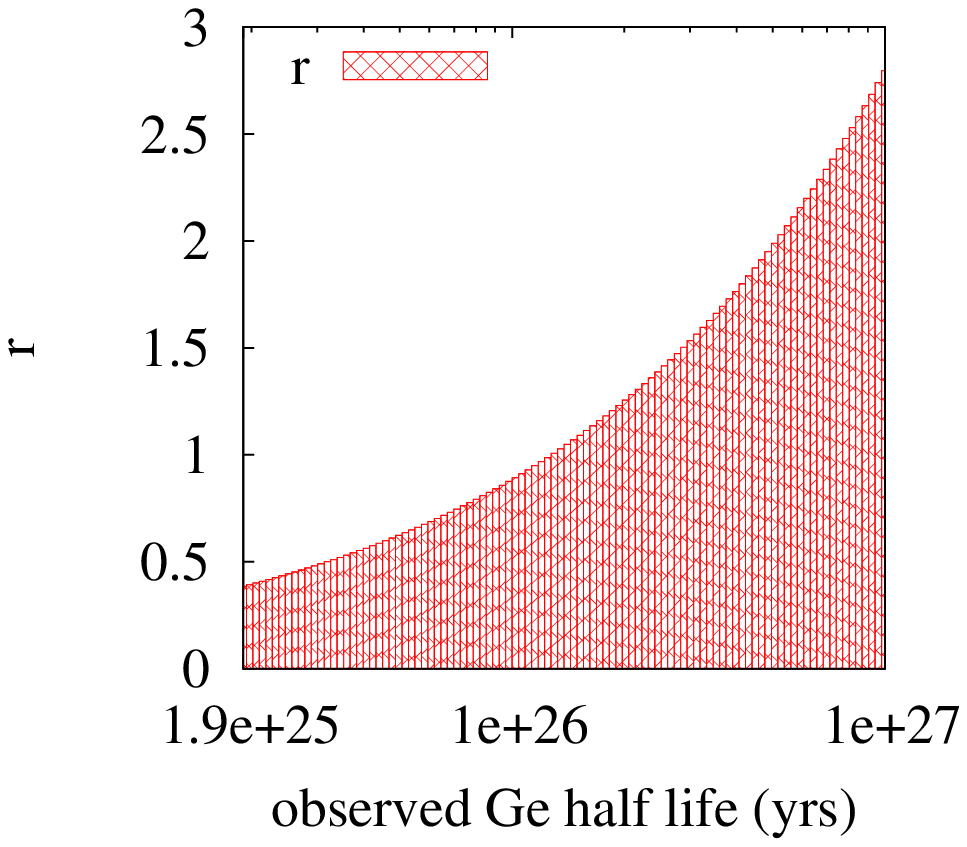}
\includegraphics[width=7cm]{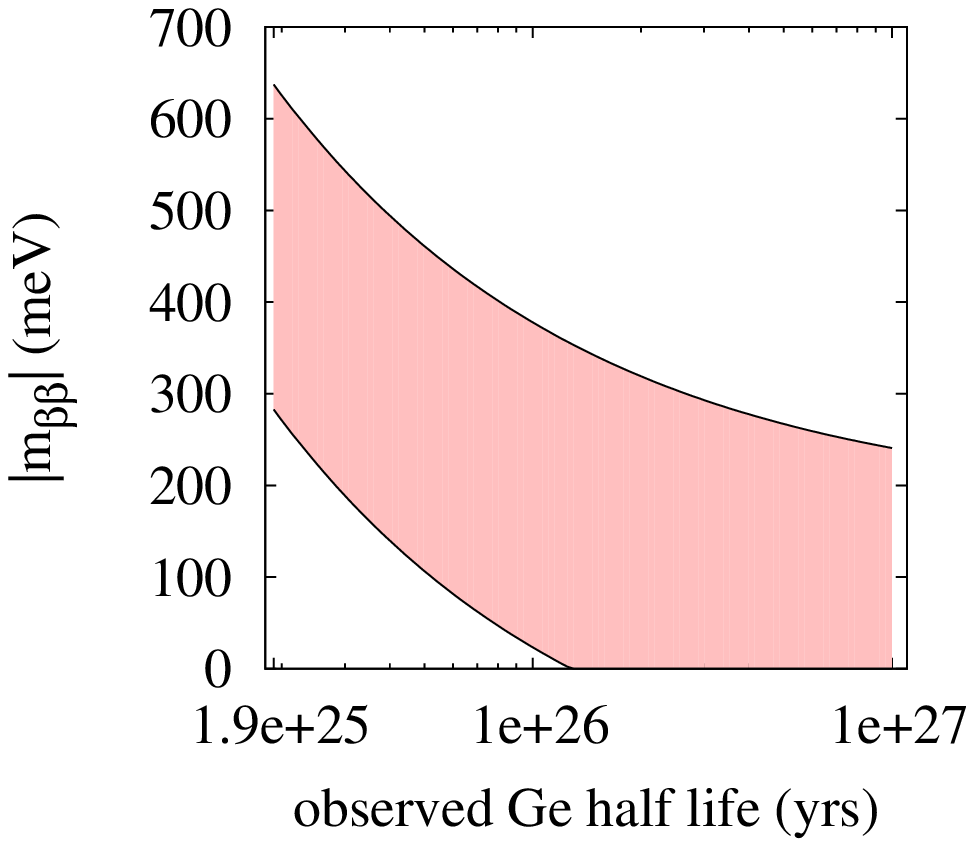}
\caption{Effect of a near-future measurement of $\bhalflife$ for $\MO=680$
  GeV, $\Mhalf=440$ GeV, given current $\bbbar$ mixing constraints. In both
  panels, the shaded regions are allowed.\label{fig:infbb}}
\end{center}
}
\end{figure}
We now consider the case where sbottom
  exchange plus other possible processes contribute to $\betadecay$, and where
  $\mee$ may not be solely due to $\lam'_{131} \lam'_{113} \neq 0$ leading to
  the process in
  fig.~\ref{fig:mee}. Thus, we consider that there may be other
  contributions to
  $\mee$,
  coming from bi-linear RPV couplings, or   $\lambda_{ijk}$ couplings,   for
  example. 
We imagine that LHC measurements are compatible with lepton-number
violating mSUGRA signals, with $\MO=680$
  GeV and $\Mhalf=440$ GeV, $\AO=0$ and $\tan \beta=10$. In practice, these
  numbers would be determined with some uncertainties, which we ignore for now
  since we are merely illustrating the point.
Then, a measurement of $\bhalflife$ in the next generation of experiments
could constrain the $\betadecay$ mechanism when combined with $\bbbar$
constraints. In order to quantify how much of the $\betadecay$ width may come
from direct processes involving sbottoms, we define
\eq
r = \left| \frac{M_{\lam'_{113}\lam'_{131}}}{M_{tot}} \right|,
\qe
where $M_{tot}$ is the total matrix element including both sbottom mediated
and $m_{\beta \beta}$-induced contributions. $r=1$ implies that the
sbottom-mediated contributions could account for all of the $\betadecay$,
whereas 
$r<1$ requires some extra component (for example from $m_{\beta \beta}$). 
$r>1$ also requires an additional component that destructively interferes with
the sbottom exchange process.  
We illustrate this in fig.~\ref{fig:infbb}, where the hatched
region in the left-hand panel shows which values of $r$ would be allowed by
current $\bbbar$ mixing constraints. We see that, for $\bhalflife<10^{26}$
years, it is not possible to explain $\betadecay$ with sbottom exchange while
simultaneously satisfying $\bbbar$ bounds. The right-hand panel shows what 
the inferred value of $m_{\beta \beta}$ may be assuming that the only
contributions are from sbottom exchange and $m_{\beta \beta}$. The range of
values comes from the possible size of the direct sbottom-mediated
contributions and the fact that the interference could be either constructive
or destructive. We again see that $\bhalflife<10^{26}$ years would imply that
the direct contribution may not account for $\betadecay$ alone. 
Theoretical uncertainties in the $\bhalflife$ predictions coming from NMEs 
would affect the inferences in fig.~\ref{fig:infbb}, widening the band in the
right-hand panel and raising the bound in the left-hand panel. 

\subsection{Single selectron production at the LHC via $\lam'_{111}$}\label{sec:lamp111}

\begin{figure}[!ht]
  \begin{center}
    {
      \begin{picture}(200,110)(0,0)
	{
          \ArrowLine(25,50)(45,70)
          \ArrowLine(25,90)(45,70)
          \DashArrowLine(85,70)(45,70){3}
          \ArrowLine(125,90)(85,70)
          \ArrowLine(105,60)(85,70) 
          \ArrowLine(105,60)(125,50) 
          \ArrowLine(165,70)(125,50)
          \DashArrowLine(125,50)(135,30){3}
          \ArrowLine(175,50)(135,30)
          \ArrowLine(175,10)(135,30)
	  \Vertex(105,60){1.5}
          \put(15,37){$d^c$}
          \put(15,87){$u$}
          \put(45,60){$\lam'_{111}$}
          \put(60,75){$\tilde{e}_L$}
          \put(128,87){$e$}
          \put(94,49){$\tilde{\chi}^0_1$}
          \put(168,67){$u$}
          \put(135,40){$\tilde{u}_L$}
          \put(125,20){$\lam'_{111}$}
          \put(178,48){$e$}
          \put(178,7){$d^c$}
	}
      \end{picture}
    }
    \caption[Production of single selectron at resonance via
      $\lam'_{111}$]{Production of a single selectron at resonance via
      $\lam'_{111}$, followed by a gauge decay into an electron with a
      neutralino LSP.  The LSP further decays into three final states via a
      virtual sparticle, leading to a same sign, di-lepton signal for the
      whole resonance production process.  
    \label{fig:singleselectronprod}
}
  \end{center}
\end{figure}
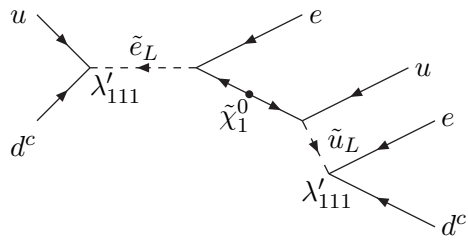
Depending on the value of $\lam'_{111}$, resonant production of a
single selectron may be observed at the LHC.  The related process of
single smuon production is studied in \cite{Dreiner:2000vf} where
like sign di-lepton signals are used because of small backgrounds.
A diagram showing like sign di-lepton production, with decay of a
selectron via a neutralino is displayed in
fig.~\ref{fig:singleselectronprod}.  Earlier studies based on
different signatures can be found in
\cite{Hewett:1998fu,Kalinowski:1997zt,Allanach:1999bf,Dreiner:1999qz,Dimopoulos:1988fr}. For previous analyses of lepton number violation 
utilizing the same sign 
di-lepton signature see e.g. \cite{Dreiner:1993ba,Kolb:2001yb,Han:2006ip,Kolb:aaa}.
A first study of single slepton production in stau LSP scenarios can
be found in \cite{Dreiner:2008rv}.  
However, to the best of our
knowledge, the discovery reach of such stau LSP scenarios at the LHC is not
available in the literature, and hence our following discussion will be
restricted to the case with a neutralino LSP. 

At low sparticle mass scales of $\sim 100$ GeV, the stringent bound
from $\betadecay$ renders single slepton production unobservable.  However
the strong dependence of this
bound on the SUSY mass scale, as shown in eq.~(\ref{eq:111directdecaylimit}),
means that it may be possible for this process
to be observed at higher SUSY mass scales.

\begin{figure}[!ht]
  \centering
  \scalebox{1.0}{
    \includegraphics[width=0.5\textwidth]{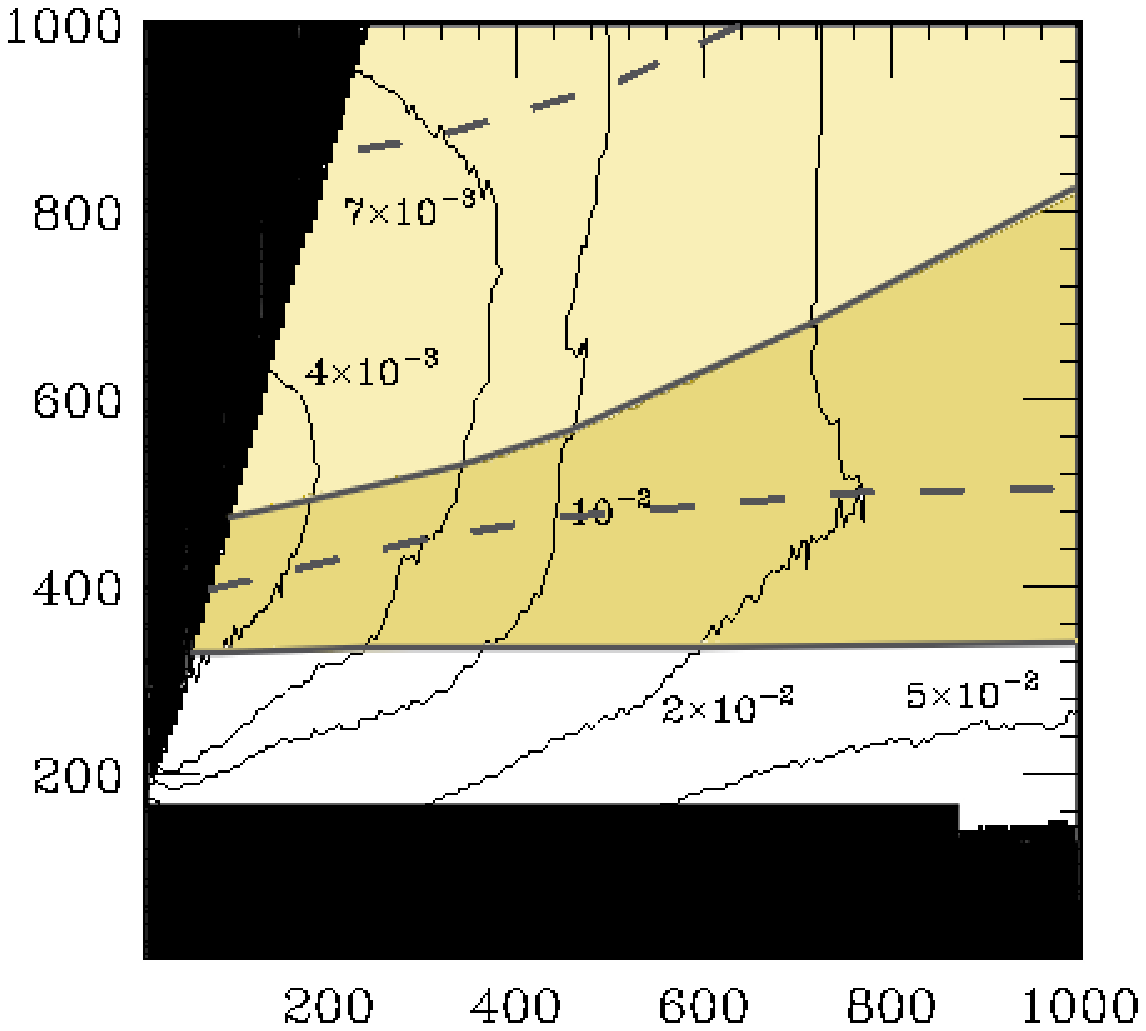}
    \put(-130,0){$\MO$/GeV}
    \put(-220,80){\rotatebox{90}{$\Mhalf$/GeV}}
  }
  \caption{mSUGRA parameter space in which
  single slepton production may be observed at the LHC for $m_{\beta
    \beta}=0$, $\tan \beta=10$,  
  $A_0=0$ and 10$\ifb$ of integrated luminosity at 14 TeV centre of mass
  energy.  
  In the top left-hand black triangle, the stau is the LSP, a case not covered
  by this analysis. The bottom black region is ruled out by direct
  search constraints. The labelled contours are extracted from
  Ref.~\cite{Dreiner:2000vf}, and show the search reach given by the labelled
  value of $\lam'_{111}$. From bottom to top, the white, dark-shaded
  and light-shaded   regions 
  show that observation of single slepton production at the $5\sigma$ level
  would imply     $\bhalflife < 1.9\cdot 10^{25} \textrm{yrs}$,  $100> \bhalflife/10^{25}  \textrm{yrs} > 1.9$ and
    $\bhalflife > 1 \times 10^{27} \textrm{yrs}$, respectively.
The upper and lower dashed curves
show where the contour between the dark-shaded and light-shaded regions would
move to if $|\mee|=0.05$ eV were included 
with constructive or destructive interference, respectively.
\label{fig:lamp111_Discovery_sleptonprod_1.9e25.eps}
}
\end{figure}

We will now compare the
discovery reach of $\lam'_{211}$ \cite{Dreiner:2000vf} at the LHC via single
smuon 
production with the bounds on $\lam'_{111}$ coming
from $\betadecay$. 
As Ref. \cite{Dreiner:2000vf} does not include detector effects, the
reach of $\lam'_{211}$ is expected to be readily applicable to
$\lam'_{111}$ without significant changes.  For this reason we shall
use the discovery reach of $\lam'_{211}$ interchangeably with that of
$\lam'_{111}$ from now on and use the results of Ref.~\cite{Dreiner:2000vf}
as an estimate of 
the 5$\sigma$-significance discovery reach for 10 fb$^{-1}$
of LHC integrated luminosity at a centre of mass energy of 14 TeV. 
Cuts were placed on the leptons: a minimum
transverse energy cut and an isolation cut in order to reduce heavy quark
backgrounds. Other SUSY processes are cut by requiring a maximum missing
transverse energy and requiring that there are at most 2 or 3 jets above a
minimum transverse momentum. We refer the interested reader to
Ref.~\cite{Dreiner:2000vf} for more details.

Fig.~\ref{fig:lamp111_Discovery_sleptonprod_1.9e25.eps} shows regions of the
$\MO-\Mhalf$ plane where single slepton production may be observed via
like-sign electrons plus two jets. 
The black regions with high $\Mhalf$ and low $\MO$ have a stau LSP.  As discussed
before, this is not included in our discussion because a detailed Monte Carlo
study is not yet available (however see \cite{Dreiner:2008rv} for an initial study).
The black regions with small $\Mhalf$ are excluded as in 
fig.~\ref{fig:113R_NMERPVmee_Faessler.eps}.
In the white region, single
slepton production by 
$\lam'_{111}$ could not be observed without violating the current bound upon
$\bhalflife$. 
The darker shaded region shows where the observation of single
slepton production at $5\sigma$ above background implies that $\betadecay$ is within the reach of the
next generation of experiments, which should be able to probe $\bhalflife < 1
\times 10^{27}\textrm{yrs}$~\cite{Avignone:2007fu,Aalseth:2004hb}. Conversely, if $\betadecay$ is discovered by the next
generation of experiments, we should expect single slepton production to be
observable and test the $\lam'_{111}$ hypothesis. We do not expect $A_0$ or
$\tan \beta$ to
affect the shape of the 
regions much, since they have a negligible effect on the
selectron mass and the couplings in the relevant Feynman diagrams. 
In the light shaded (upper) region, a 5$\sigma$ single slepton discovery at
the LHC implies that
the next generation of experiments would not be able to observe $\betadecay$.
Conversely, if $\betadecay$ is within reach of the next generation of
experiments, the LHC would see single slepton production signal in this region
at greater than $5 \sigma$ significance.

If $\betadecay$ is marginally observed at the limit of the next round
of $\betadecay$ experiments, it is possible that a contribution from $\mee$
could contribute significantly (if it originated from a different coupling to $\lam'_{111}$).  For $\Ge$, a reach of $\bhalflife \sim
10^{27}$yrs implies that an inverted or quasi-degenerate neutrino mass
spectrum may contribute to $\betadecay$ at an observable
level.
Whether it interferes with the direct
contribution amplitude constructively or not will affect
the potential observability at the LHC, and should be taken into account.  
We shall assume that
$|\mee| = \sqrt{\atmos} \sim 0.05\textrm{~eV}$, as implied by neutrino
oscillation data.
We
show in fig.~\ref{fig:lamp111_Discovery_sleptonprod_1.9e25.eps} 
where the upper edge of the darker region would move to 
and constructive (upper dashed curve) or destructive (lower dashed curve)
interference between the $\mee$ contribution and the direct contributions.
We
see that for the case of constructive interference in
fig.~\ref{fig:lamp111_Discovery_sleptonprod_1.9e25.eps}, discovery of single
slepton production becomes very difficult with $10\,\ifb$ of data.  On
the other hand, the situation improves substantially if destructive
interference occurs instead.  
To see this, we note that for a fixed $\bhalflife$, the introduction of a
non-vanishing $\mee$ which interferes destructively with $M_{\lam'_{111}}$ would
imply an 
increase in the direct contributions.  This would require an increase in 
$\lam'_{111}$, or a decrease in the mass of the SUSY spectrum, or both, all of
which lead to an increase in single slepton production rate. 
We see that single
selectron production may be observed for $\Mhalf \gtrsim 500$ GeV.  If
the neutrino mass spectrum is instead normal hierarchical, the $\mee$
diagram will be sub-dominant for the half life discussed above.  This
situation may be approximated by setting $\mee=0$.

\begin{figure}[!ht]
  \centering
  \scalebox{1.0}{
    \includegraphics[width=0.7\textwidth]{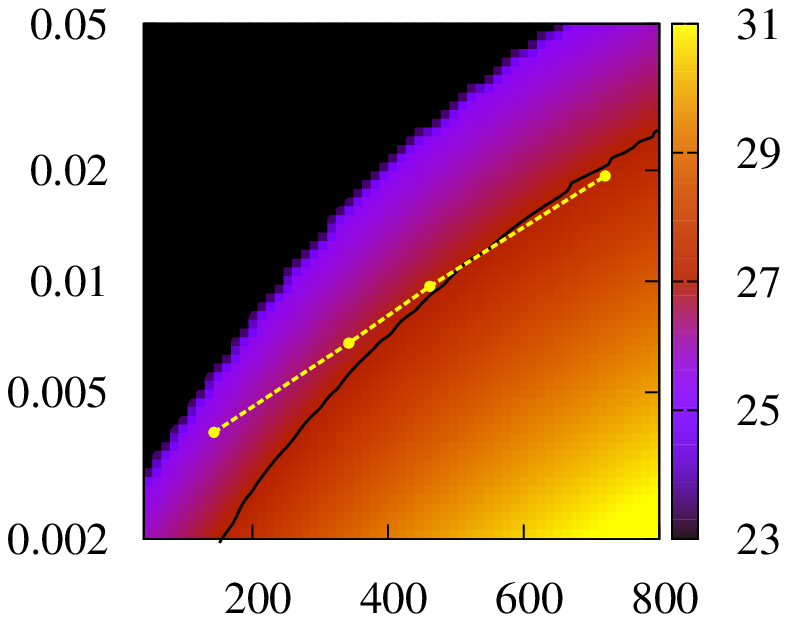}
    \put(-170,15){$\MO$/GeV}
    \put(-263,100){\rotatebox{90}{$\lam'_{111}$}}
    \put(-110,182){\small $\textrm{log}_{10}\bhalflife$}
  }
  \caption{
Comparison of $\bhalflife$ and single slepton discovery reach as a
    function of $\lam'_{111}$ along the mSUGRA slope $M_{1/2}=300\mbox{~GeV}+0.6M_{0}$,
    with $\AO=0$, $\tanb=10$ and $\textrm{sgn}\mu=1$.  The black region on the
    top left corner is ruled out by $\betadecay$.  The region above the solid
    black line is accessible in near future $\betadecay$ experiments, whereas
    the light dotted line shows the lower limit of $\lam'_{111}$ for single
    slepton production to be discoverable at the
    LHC. \label{fig:1027}}
\end{figure}
We show in fig.~\ref{fig:1027} the variation of the discovery reach of $\lam'_{111}$
with $M_{0}$ along the line $M_{1/2}=300 \mbox{~GeV} + 0.6M_{0}$.  Above the
dotted light line, single slepton production  
will be observed at the LHC\@. We see from the figure that for nearly all of the
parameter space where $\betadecay$ can be measured by the next generation of
experiments,  the LHC would provide a confirmation of the supersymmetric
origin of the signal by observing single slepton production at the 5$\sigma$
level. 

\begin{figure}[!ht]{
\begin{center}
\includegraphics[width=7cm]{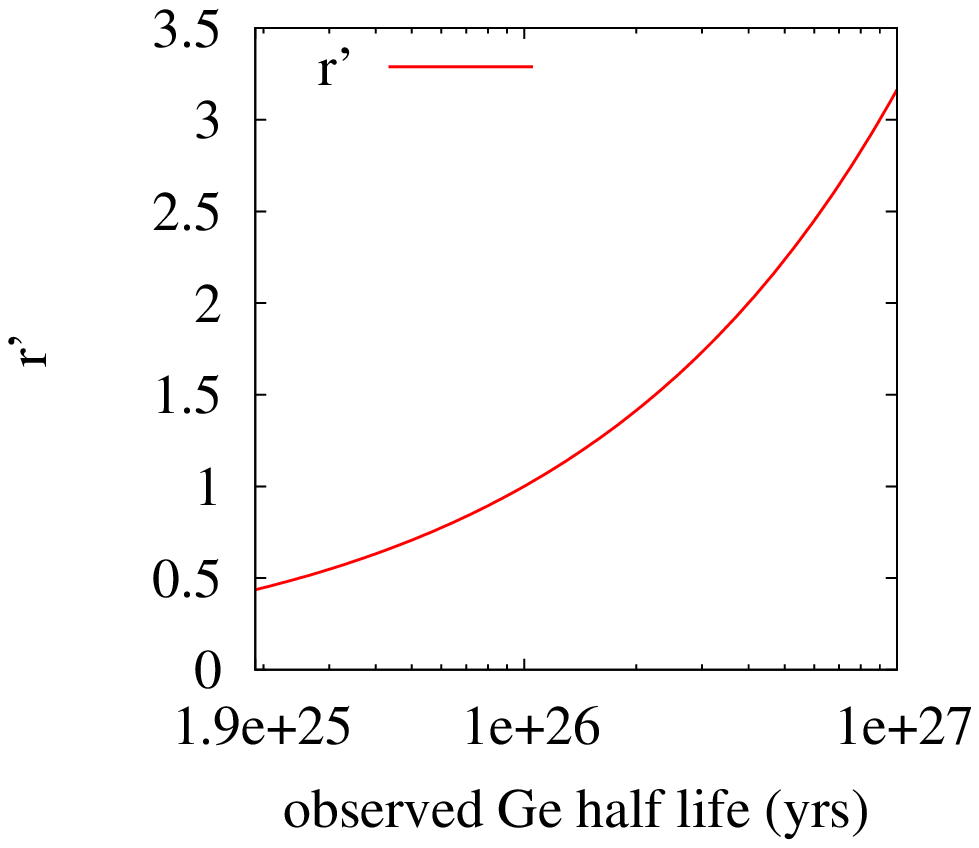}
\includegraphics[width=7cm]{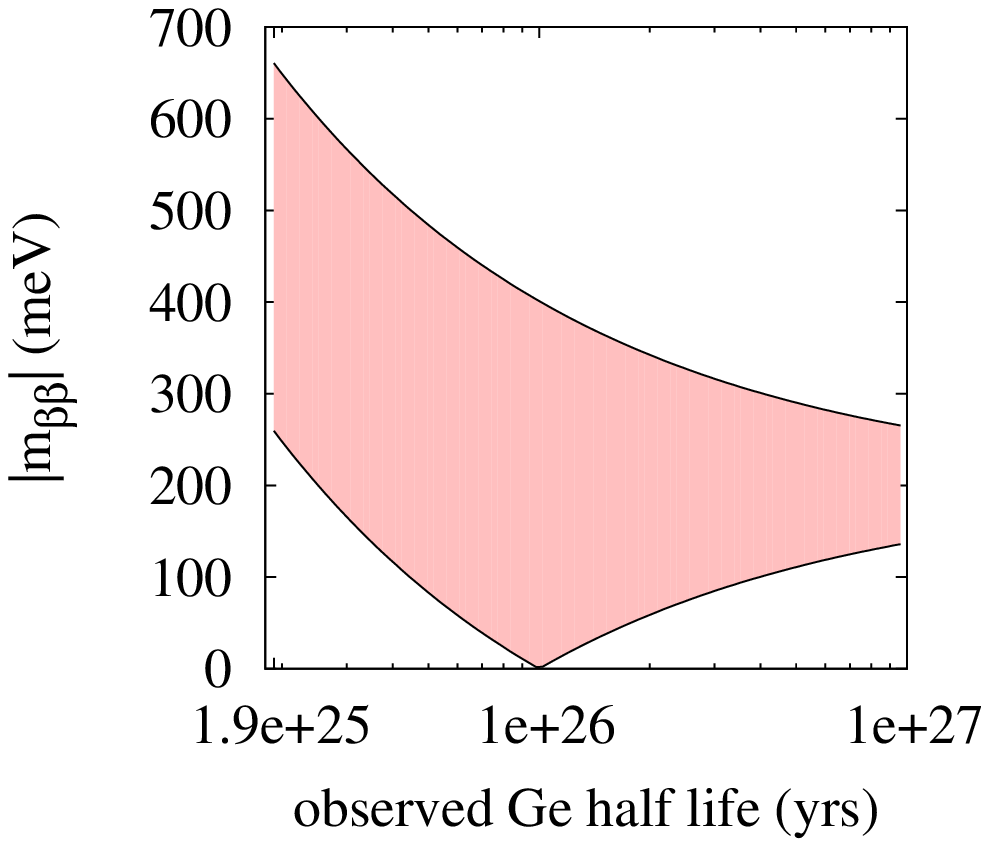}
\caption{Effect of a near-future measurement of $\bhalflife$ for $\MO=680$
  GeV, $\Mhalf=440$ GeV, given a 5$\sigma$ observation of single slepton
  production at the LHC in 10 fb$^{-1}$ at 14 TeV centre of mass energy.
  In the right-hand panel, the shaded region would be allowed. \label{fig:insel}}
\end{center}
}
\end{figure}
We now ask the question: if single slepton production were observed at the
LHC, and a measurement of $\bhalflife$ were made, what could be divined about
the relative contributions between direct or Majorana neutrino-induced
$\betadecay$? We define
\eq
r' = \left| \frac{M_{\lam'_{111}}}{M_{tot}} \right|,
\qe
where $M_{tot}$ is the total matrix element including both sbottom mediated
and $m_{\beta \beta}$-induced contributions. At our parameter test point of
$\MO=680$   GeV, $\Mhalf=440$ GeV, $\AO=0$ and $\tan \beta=10$, a measurement
of the single slepton production cross-section could be used to infer a value
of $\lam'_{111}$, which would then imply a value of $r'$. Here, we assume that 
the cross-section was just at the 5$\sigma$ level above background and show
which $r'$ could be inferred from the measured value of $\bhalflife$ in
the left-hand panel of fig.~\ref{fig:insel}. 
The figure shows a definite prediction for $r'$, which is 1 at a half-life of
$10^{26}$ years, where $\betadecay$ could come entirely from the
direct contribution. 
In general, $r'>1$ requires destructive interference between 
the $\lam'_{111}$ contribution and another contribution (either
$\lam'_{131}\lam'_{113}$ or $\mee$). Assuming the relative phases between different
contributions are real, in the destructive interference case there are two solutions
to $|M_{\lam'_{111}}-M_{\mbox{other}}|=|M_{tot}|$, where $|M_{tot}|$ 
is a prescribed constant derived from $\bhalflife$ and
eq.~(\ref{eq:halflifeformula}).  
We note that
NME uncertainties, which are neglected here, would turn the definite
prediction into a band of possible predictions. 
The $r'$ prediction would acquire error bands, at the level
of $\sim {\mathcal O}(20)\%$, from measurement errors in the SUSY spectrum,
cross-section and NME uncertainties. A further dedicated study including
simulations of the experiments is required in order to quantify these latter
errors more exactly. 

The right-hand panel of fig.~\ref{fig:insel} shows
correspondingly what may be deduced about $m_{\beta \beta}$, also neglecting
measurement and NME errors, but taking into 
account two possible contributions to the matrix element: $M_{\nu}$ and
$M_{\lam'_{111}}$. If instead we assumed that $\betadecay$ was due entirely to 
the Majorana neutrino induced contribution,  then one
deduces $\mee$ from $\bhalflife$ as in fig.~\ref{fig:meein}. We see that the
definite prediction widens to a band, due to the relative complex phase 
between the two contributions to the matrix element. 
Any inferred value $\mee\gsim 10$ meV corresponds to a
non-hierarchical pattern (i.e.\ inverted or quasi-degenerate) of neutrino
masses, and so one would infer that the 
hierarchy is non-hierarchical for $\bhalflife \gsim 10^{27}$ years. 
We see that, for this parameter space point, if $\bhalflife \approx 10^{26}$
years,  
the observation of single slepton would imply that a normal hierarchy is 
still viable, although still one could not tell without
additional measurements if the neutrino mass hierarchy were inverted or
hierarchical. On the other hand, for 
$\bhalflife > 1.1 \times 10^{26}$ years or 
$\bhalflife < 0.9 \times 10^{26}$ years, neutrino masses have the
non-hierarchical pattern. It remains to be seen how large this window remains
after measurement uncertainties are calculated. 
In the limit where the contributions from 
$\mee$ and
$M_{\lam'_{111}}$ are real, a given
$\bhalflife$ would result in two-fold possible predictions: either the top or
the bottom of the bands in the right-hand panel of
fig.~\ref{fig:insel}. Theoretical errors originating from the NME calcuations
would widen the band further. 

\section{Discussion and summary \label{sec:disc}}
In this paper, we have discussed the interplay between a number of observables
in RPV SUSY: $\betadecay$, $\bbbar$ mixing and first generation single slepton
production at the LHC.  We saw that, while it is difficult to infer
unambiguously the presence of RPV effects from individual observables, further
insight could be gained if these information are analysed simultaneously.
Bounds from $\bbbar$ mixing and single slepton production could constrain the
extent to which $\lam'_{113}\lam'_{131}$ and $\lam'_{111}\lam'_{111}$
operators may contribute to $\betadecay$.  They provide extra handles to
complement other strategies for divining the underlying $\betadecay$
mechanisms, for 
example by comparing life times of different nuclei \cite{Deppisch:2006hb}.

There are a couple of caveats one should bear in mind however.  Firstly, our
numerical inference between collider observables and $\betadecay$ depends on the
relation between the masses of the
SUSY particles that mediate the different processes. Constraints on these
masses from experiment are required: 
ideally directly, but otherwise because some other simple theoretical SUSY
breaking ansatz such as mSUGRA fits LHC data well. 
Secondly, the estimated $\bhalflife$ is subject to uncertainties in the NME.
This is 
important particularly when using $\bbbar$ data because the sensitivity of the
next round of $\Ge$ $\betadecay$ experiments is not expected to go far beyond
the $10^{27}$ yrs level.  The bounds from $\betadecay$ and $\bbbar$ are
comparable in regions with 0.1--1 
TeV scale SUSY breaking, with the latter being slightly more stringent in the
parameter space we explored.  An increase in the NME value used might change
this observation. 

At present, $\bbbar$ mixing implies only an upper bound on the possible 
contribution of heavy sbottom exchange to any measurement of $\bhalflife$,
given information on the SUSY spectrum from the LHC. 
Sbottom exchange could be solely responsible for 
$\bhalflife(\Ge)$ of $\sim 10^{26}\textrm{yrs}$, which
is potentially observable in the near future.
However, the measured
value of $\bhalflife$ could still require there to be a non-zero Majorana
neutrino contribution, depending upon its value.
A much more informative two-sided bound on the sbottom exchange contribution
would result if a future measurement of $\bbbar$ mixing required an extra component
from outside the Standard Model. 

We point out that, contrary to previous expectations, a
same sign di-lepton signal from single selectron production via
gauge decay at the LHC could be observed.  This is despite the stringent bound
from $\betadecay$, because the constraint from
$\betadecay$ on $\lam'_{111}$ relaxes rapidly as the SUSY scale increases.
We have shown that if a direct
contribution via $\lam'_{111}$ is the dominant $\betadecay$ mechanism and
$\betadecay$ of $\Ge$ is just beyond the current reach, 
there is a good chance of observing single selectron production at the
LHC. Such a scenario is not ruled
out by current $\betadecay$ bounds for a heavy enough SUSY spectrum. 
Knowledge of the SUSY spectrum can be combined with $\betadecay$ and
single-selectron data to bound the $\lam'_{111}$ contribution to $\betadecay$.
Thus, evidence for other contributions may be obtained (for example from
Majorana neutrino exchange) and the size of the other contributions bounded. 
Under the hypothesis that only the $\lam'_{111}$ and $\mee$ contributions are
significant, $\mee$ may be deduced and information about the neutrino mass
spectrum is thus obtained. For some ranges of $\bhalflife$, this could 
settle the question of whether the neutrino spectrum were
hierarchical or not without the inclusion of other observables. 

It will be an interesting exercise in future work to examine a particular
point in parameter space with a dedicated LHC simulation study in order to
quantify the errors obtained on the inferred contributions to $\betadecay$ in
a combined fit.

\section*{Acknowledgments}
This work has been partially supported by STFC.  We thank the
Cambridge SUSY working group, G. Hiller, M. Hirsch and R. Mohapatra for
useful conversations.  CHK is funded by a Hutchison Whampoa Dorothy Hodgkin
Postgraduate Award. HP was partially funded by the EU project ILIAS N6 WP1.
BCA and CHK also thank the Technische Universit\"at
Dortmund, and HP thanks the University of Cambridge for hospitality offered
while part of this work was carried out. Some of the hospitality extended to
BCA arose from a Gambrinus Fellowship awarded by the Technische Universit\"at
Dortmund.

\appendix
\section{Parton Level Contributions to $\betadecay$ \label{sec:part}}
\subsection{Light Majorana neutrino exchange: $\betadecay$ via $\mee$}
The effective Lagrangian after integrating out the $W$ gauge boson,
and the $\Delta L_e=2$ Lagrangian with a virtual Majorana neutrino are
\eqa\label{eq:SMmee}
\lag^{eff}_{EW}(x) &=& -\frac{G_F}{\sqrt{2}} \Big[ \bar{e}\gamma^{\mu}(1-\gamma_5)\nu\,\,\bar{u}_y\gamma_{\mu}(1-\gamma_5)d^y \Big], \nonumber \\
\lag^{eff,\,\Delta L_e=2}_{EW}(x) &=& \frac{G_F^2}{2} \Big[ \bar{e}\gamma_{\mu}(1-\gamma_5)\frac{\mee}{q^2}\gamma_{\nu}e^c \Big] \Big[ J^{\mu}_{V-A}J^{\nu}_{V-A} \Big],
\qea
respectively.
$y$ is a colour index, and in the second line of eq.~(\ref{eq:SMmee}),
$J^{\mu}_{V-A} = \bar{u}_y\gamma^{\mu}(1-\gamma_5)d^y$.  A Feynman diagram
depicting this process is displayed in fig.~\ref{fig:mee0vbb}. 

\subsection{Heavy sbottom exchange: $\betadecay$ via $\lam'_{113}\lam'_{131}$}
In the basis where both the down-type quark and the charged lepton mass matrices are diagonal, the coupling product $\lam'_{113}\lam'_{131}$ leads to an effective Lagrangian involving exchange of one SUSY particle of the form
\eqa \label{eq:lamp113Lag}
\lag^{eff}_{\lam'_{113}\lam'_{131}}(x) &=& \frac{G_{F}}{8\sqrt{2}}\eta^n(U^*_{PMNS})_{ni}\Big[\frac{1}{2}(\bar{\nu}_{i}\sig^{\mu\nu}(1+\gamma_5)e^{c})(\bar{u}_{y}\sig_{\mu\nu}(1+\gamma_5)d^{y}) \nonumber  \\
&&\qquad\quad\qquad\quad\qquad+2(\bar{\nu}_{i}(1+\gamma_5)e^{c})(\bar{u}_{z}(1+\gamma_5)d^{z})\Big],
\qea
where \cite{Babu:1995vh}\footnote{The first term in eq.~(\ref{eq:lamp113Lag})
  differs from \cite{Pas:1998nn,Faessler:2007nz} and also the preprint version
  of \cite{Babu:1995vh} by a factor of $\frac{1}{2}$.  Our check agrees with
  the published version of the latter reference.\label{footnote:factorhalf}} 
\eqa \label{eq:eta}
\eta^n &=& \sum_{k}\frac{\lam'_{1mk}\lam'_{nk1}(U_{CKM})_{1m}}{2\sqrt{2}G_F}\textrm{sin}2\theta_{\tilde{d}_{k}}\Big(\frac{1}{m^2_{\tilde{d}_{k(1)}}}-\frac{1}{m^2_{\tilde{d}_{k(2)}}}\Big) \nonumber \\
&\simeq& -\frac{\lam'_{113}\lam'_{n31}}{\sqrt{2}G_F}\Big(\frac{m^2_{\tilde{b}_{LR}}}{m^2_{\tilde{b}_{LL}}m^2_{\tilde{b}_{RR}}-m^4_{\tilde{b}_{LR}}}\Big).
\qea

In eq.~(\ref{eq:lamp113Lag}), $U_{PMNS}$ is the 3 $\times$ 3 unitary 
PMNS neutrino mixing matrix \cite{Pontecorvo:1967fh,Maki:1962mu}, and
$\sig^{\mu\nu}=\frac{i}{2}[\gamma^{\mu},\gamma^{\nu}]$.  
In eq.~(\ref{eq:eta}), $U_{CKM}$ is the CKM matrix and
$m^2_{\tilde{d}_{k(LL)}}$, $m^2_{\tilde{d}_{k(LR)}}$ and
$m^2_{\tilde{d}_{k(RR)}}$ denote entries in the k-th generation down type
squark mass squared matrix.  In particular, $m_{\tilde{b}_{1}}$ and
$m_{\tilde{b}_{2}}$ denote the 2 sbottom mass eigenvalues, and
$\theta_{\tilde{b}}$ is the sbottom left-right mixing angle.  The relations
between the mixing angle and the entries in the mass and flavour basis sbottom
mass matrices follow those in \softsusy\cite{Allanach:2001kg}.  

The complete $\betadecay$ matrix element with a
leptonic current coupled to a quark current via a $W$ boson does not contain a
neutrino mass insertion, and hence is not suppressed by the light neutrino
mass scale.  The $\Delta L_e=2$ Lagrangian is given by 
\eqa\label{eq:lamp113131SM0vbb}
\lag^{eff,\,\Delta L_e=2}_{EW+\lam'_{113}\lam'_{131}}(x) &=& \frac{G^2_F}{2} \eta^1 \Big[ \frac{1}{2} \Big(\bar{e} \gamma_{\rho} (1-\gamma_5) \frac{1}{\not\!q} e^c\Big) J^{\rho}_{V-A}J_{PS} \nonumber \\
&&\qquad\quad + \frac{1}{8} \Big(\bar{e} \gamma_{\rho}(1-\gamma_5)\frac{1}{\not\!q} \sigma_{\mu\nu} e^c\Big) J^{\rho}_{V-A}J^{\mu\nu}_T \Big],
\qea
where $J_{PS}=\bar{u}_y(1+\gamma_5)d^y$ and $J^{\mu\nu}_T =
\bar{u}_y\sigma^{\mu\nu}(1+\gamma_5)d^y$ are the pseudo-scalar and tensor
currents respectively.  A Feynman diagram depicting this process is displayed
in fig.~\ref{fig:lamp1130vbb}. 
The matrix element is \cite{Pas:1998nn,Faessler:2007nz}
\eq \label{eq:matrixelement113}
M_{\lam'_{113}\lam'_{131}} = \eta^1 (M^{2N}_{\tilde{q}} +
M^{\pi}_{\tilde{q}}),  
\qe
$\eta^1$ is
defined as in eq.~(\ref{eq:eta}) with $n=1$, and $M^{2N}$ and $M^{\pi}$ denote
the 2 nucleon mode and pion mode contributions and will be detailed in
Appendix~\ref{nmes}. 

\subsection{Sparticle exchange: $\betadecay$ via $\lam'_{111}\lam'_{111}$}
Following the notation of \cite{Hirsch:1995ek}, the effective Lagrangian with $\lam'_{111}$ in the direct RPV $\betadecay$ process involving exchange of two SUSY particles is given by
\eqa
\lag^{eff,\,\Delta L_e=2}_{\lam'_{111}\lam'_{111}}(x) &=& \frac{G^2_{F}}{2}m_p^{-1}[\bar{e}(1+\gamma_5)e^c] \nonumber \\
&&\times \Big[(\eta_{\tilde{g}} + \eta_{\chi})(J_{PS}J_{PS}-\frac{1}{4}J^{\mu\nu}_T J_{T\mu\nu}) + (\eta_{\chi\tilde{e}}+\eta'_{\tilde{g}}+\eta_{\chi\tilde{f}})J_{PS}J_{PS}\Big],
\qea
where the RPV coefficients are defined to be
\eqa \label{eq:lamp111eta}
\eta_{\tilde{g}} &=& \frac{\pi\alpha_s}{6} \frac{\lam'^2_{111}}{G^2_F} \frac{m_p}{m_{\tilde{g}}} (\frac{1}{m^4_{\tilde{u}_L}} + \frac{1}{m^4_{\tilde{d}_R}} - \frac{1}{2m^2_{\tilde{u}_L}m^2_{\tilde{d}_R}}), \nonumber \\
\eta_{\chi} &=& \frac{\pi\alpha_2}{2} \frac{\lam'^2_{111}}{G^2_F} \sum_{i=1}^4 \frac{m_p}{m_{\chi_i}} (\frac{\epsilon^2_{L_i}(u)}{m^4_{\tilde{u}_L}} + \frac{\epsilon^2_{R_i}(d)}{m^4_{\tilde{d}_R}} - \frac{\epsilon_{L_i}(u)\epsilon_{R_i}(d)}{m^2_{\tilde{u}_L}m^2_{\tilde{d}_R}}),  \nonumber \\
\eta_{\chi\tilde{e}} &=& 2\pi\alpha_2 \frac{\lam'^2_{111}}{G^2_F} \sum_{i=1}^4 \frac{m_p}{m_{\chi_i}} (\frac{\epsilon^2_{L_i}(e)}{m^4_{\tilde{e}_L}}),  \nonumber \\
\eta'_{\tilde{g}} &=& \frac{2\pi\alpha_s}{3} \frac{\lam'^2_{111}}{G^2_F} \frac{m_p}{m_{\tilde{g}}}(\frac{1}{m^2_{\tilde{u}_L}m^2_{\tilde{d}_R}}),  \nonumber \\
\eta_{\chi\tilde{f}} &=& \pi\alpha_2 \frac{\lam'^2_{111}}{G^2_F} \sum_{i=1}^4 \frac{m_p}{m_{\chi_i}} (\frac{\epsilon_{L_i}(u)\epsilon_{R_i}(d)}{m^2_{\tilde{u}_L}m^2_{\tilde{d}_R}} - \frac{\epsilon_{L_i}(u)\epsilon_{L_i}(e)}{m^2_{\tilde{u}_L}m^2_{\tilde{e}_L}} - \frac{\epsilon_{L_i}(e)\epsilon_{R_i}(d)}{m^2_{\tilde{e}_L}m^2_{\tilde{d}_R}}),
\qea
and again we follow the notation of \cite{Hirsch:1995ek}.  The $\epsilon$'s denote rotations between mass and gauge eigenbasis in the gaugino-fermion-sfermion vertices.  To facilitate comparisons with the literature, we display only the first generation sparticle contribution above but include contributions from all three generations in the numerical calculations.  The relevant Feynman diagrams are displayed in fig.~\ref{fig:lamp1110vbb}.  
Note that our expressions above are different from those presented in
\cite{Hirsch:1995ek}, \cite{Faessler:1998qv} and \cite{Faessler:1999zg}.
Reference \cite{Hirsch:1995ek} made an approximation when extracting the
colour singlet currents from the Feynman diagrams.  Also, the expressions for
$\eta_{\chi}$ and $\eta_{\chi\tilde{f}}$ (following the convention of
\cite{Hirsch:1995ek}) are different, with the discrepancy coming from the
colour flows of fig.~\ref{fig:lamp111e} and fig.~\ref{fig:lamp111f}.
References \cite{Faessler:1998qv,Faessler:1999zg} took proper account of the
colour singlet extraction, which we have checked independently.  However
$\eta_{\chi}$ and $\eta_{\chi\tilde{f}}$ remained the same as in
\cite{Hirsch:1995ek}.  
In the parameter space we explore,
the numerical differences induced by such changes in
the $\eta$s are small compared to the uncertainty in the nuclear
matrix elements.

$M_{\lam'_{111}}$ is given by~\cite{Faessler:1998qv,Hirsch:1995ek}
\eqa \label{eq:matrixelement111}
M_{\lam'_{111}} &=& (\eta_{\tilde{g}} + \eta_{\chi}) M^{2N}_{\tilde{g}} + (\eta_{\chi\tilde{e}} + \eta'_{\tilde{g}} + \eta_{\chi\tilde{f}}) M^{2N}_{\tilde{f}} \nonumber \\
&&+ \frac{3}{8}\Big((\eta_{\tilde{g}} + \eta_{\chi}) + \frac{5}{3}(\eta_{\tilde{g}} + \eta_{\chi} + \eta_{\chi\tilde{e}} + \eta'_{\tilde{g}} + \eta_{\chi\tilde{f}})\Big) (\frac{4}{3}M^{1\pi} + M^{2\pi}).
\qea
The
$\eta$s correspond to those defined in eq.~(\ref{eq:lamp111eta}), and
$M^{2N}_{\tilde{g},\tilde{f}}$, $M^{1\pi}$ and $M^{2\pi}$ denote the 2
nucleon, 1 pion and 2 pion exchange modes respectively.     They are detailed
in Appendix~\ref{nmes}.

\section{Nuclear Matrix Elements \label{nmes}}

\begin{table}[ht!]
  \centering
  \begin{tabular}{|ccll|}
    \hline
    channel & NME($\Ge$) & value & Ref. \\
    \hline
    $\lam'_{111}\lam'_{111}$ &$M^{2N}_{\tilde{g}}$   &283&   \cite{Hirsch:1995ek}\\
    &$M^{2N}_{\tilde{f}}$   &13.2&  \cite{Hirsch:1995ek}\\
    &$M^{1\pi}$            &-18.2& \cite{Faessler:1998qv}\\
    &$M^{2\pi}$            &-601&  \cite{Faessler:1998qv}\\
    \hline
    $\lam'_{113}\lam'_{131}$ &$M^{2N}_{\tilde{q}}$   &-0.9&   \cite{Faessler:2007nz}\\
    &$M^{\pi}_{\tilde{q}}$   &604&   \cite{Faessler:2007nz}\\
    \hline
    $\mee$ &$M_{\nu}$             &2.8&   \cite{Simkovic:1999re}\\
    \hline
  \end{tabular}
  \caption[Nuclear matrix elements of $\Ge$ adopted for our analysis]{Nuclear matrix elements of $\Ge$ used.  The value of the 2 nucleon mode contribution to $M_{\lam'_{113}\lam'_{131}}$ includes a factor of $\frac{1}{2}$ discussed in footnote \ref{footnote:factorhalf}.  For model details of the NME calculations, we refer readers to the literature.}\label{tab:NMEinput}
\end{table}
As we are primarily
interested in collider effects due to couplings from the direct contributions,
we adopt numerical values collected from various sources in the literature for
our discussions here. 
These parameters are displayed in table~\ref{tab:NMEinput}.


\end{document}

In this paper we concentrate on
constraints from heavy flavour physics and the Large Hadron Collider (LHC) that
provide indirect information on possible $\betadecay$ mechanisms in the RPV MSSM.
First, the $\betadecay$ bound on the direct contribution mediated by the coupling
product $\lam'_{113}\lam'_{131}$ depends on the two sbottom masses.  The same
coupling product is also constrained by $\bbbar$ mixing  \cite{Choudhury:1996ia,Barbieri:1997zn,Pas:1998nn}, with the bound
being dependent on the electron sneutrino mass.  Knowledge of the masses of the
electron sneutrino and that of the sbottoms will therefore allow us
to infer whether this direct contribution mechanism could be the dominant source of
$\betadecay$.  Second, the coupling $\lam'_{111}$ can lead to
single selectron production at hadron colliders (see \cite{Dreiner:2000vf,Hewett:1998fu,Kalinowski:1997zt,Allanach:1999bf,Dreiner:1999qz,Dimopoulos:1988fr,Dreiner:2008rv} for an analysis on the related process of single smuon production and the associated bound).  Even though there exist stringent bounds on the product $\lam'_{111}\lam'_{111}$ \cite{Mohapatra:1986su,Hirsch:1995ek,Allanach:1999ic,Faessler:1998qv} from non-observation 
of $\betadecay$ for a low SUSY mass scale of around 100 GeV, its strong 
dependence on the SUSY mass scale means that this bound relaxes rapidly as 
one raises the SUSY scale, making single selectron production at the LHC 
potentially observable for heavier mass spectra.

Because of the presence of direct contributions in the presence of R-parity violating operators, one cannot infer unambiguously the value of $\mee$ from the lower bound on $\betadecay$ half-life $\bhalflife$.  
Nevertheless, since the RPV couplings in turn generate neutrino masses,
knowledge of the magnitude of the relevant RPV coupling could 
still allow us to obtain information on $\mee$ and hence the possible 
neutrino mass spectrum.  

\bibitem{Barbieri:1997zn}
  R.~Barbieri, A.~Strumia and Z.~Berezhiani,
  Phys.\ Lett.\  B {\bf 407} (1997) 250
  [arXiv:hep-ph/9704275].